\documentclass[%
 preprintnumbers,
 amsmath,amssymb,
 aps,
]{revtex4-2}

\usepackage{graphicx}
\usepackage{dcolumn}
\usepackage{bm}

\usepackage[a4paper, total={6in, 9in}]{geometry}
\usepackage{graphicx}
\usepackage{amssymb}
\usepackage{amsthm}
\usepackage{bbm}
\usepackage{bm}
\usepackage{amsmath}
\usepackage{mathrsfs}
\usepackage{amssymb}
\usepackage{exscale}
\usepackage{easybmat}
\usepackage{xcolor}
\usepackage{comment}
\usepackage[normalem]{ulem}
\usepackage{booktabs}
\usepackage{multirow}
\usepackage{siunitx}

\newcommand{\ii}{\mathrm{i}}
\newcommand{\ee}{\mathrm{e}}

\begin{document}


\title{Complex Spatiotemporal Modulations and Non-Hermitian Degeneracies in $\mathcal{PT}$-Symmetric Phononic Materials}

\author{M. Moghaddaszadeh$^{1,2}$}
\author{M. A. Attarzadeh$^2$}
\author{A. Aref$^1$}%
\author{M. Nouh$^2$}
 \email{Corresponding author: mnouh@buffalo.edu}
\affiliation{%
$^1$Department of Civil, Structural and Environmental Engineering, University at Buffalo (SUNY), Buffalo, NY 14260-4300, USA\\
$^2$Department of Mechanical and Aerospace Engineering, University at Buffalo (SUNY), Buffalo, NY 14260-4400, USA
}%


\begin{abstract}
Unraveling real eigenfrequencies in non-Hermitian $\mathcal{PT}$-symmetric Hamiltonians has opened new avenues in quantum physics, photonics, and most recently, phononics. However, the existing literature squarely focuses on exploiting such systems in the context of scattering profiles (i.e., transmission and reflection) at the boundaries of a modulated waveguide, rather than the rich dynamics of the non-Hermitian medium itself. In this work, we investigate the wave propagation behavior of a one-dimensional non-Hermitian elastic medium with a universal complex stiffness modulation which encompasses a static term in addition to \textit{real} and \textit{imaginary} harmonic variations in both space and time. Using plane wave expansion, we conduct a comprehensive dispersion analysis for a wide set of sub-scenarios to quantify the onset of complex conjugate eigenfrequencies, and set forth the existence conditions for gaps which emerge along the wavenumber space. Upon defining the hierarchy and examining the asymmetry of these wavenumber gaps, we show that both the position with respect to the wavenumber axis and the imaginary component of the oscillatory frequency largely depend on the modulation type and gap order. Finally, we demonstrate the coalescence of multiple Bloch-wave modes at the emergent exceptional points where significant direction-dependent amplification can be realized by triggering specific harmonics through a process which is detailed herein. 
\end{abstract}

\maketitle


\section{Introduction \label{Sec:I}}

Modulating the properties of elastic media, whether in lattice or continuum form has been a growing topic of interest in the field of phononic materials \cite{nassar2020nonreciprocity}. In its simplest form, a phononic crystal represents a composite periodic structure of two or more materials with different properties, which give rise to frequency band gaps as a result of Bragg scattering and destructive interferences \cite{al2019dispersion}. This discrete periodic alternation of materials represents the most basic form of spatial material modulation. However, phononic crystals with harmonically varying stiffness profiles represent a different form of continuous modulation where the material change along the spatial coordinate follows a prescribed function, e.g., a sinusoid \cite{su2012influence,ansari2017analyzing}. Lately, the ability to achieve wave filtering using temporal modulations of material properties via real time-periodic functions has been demonstrated \cite{trainiti2019time, Vila2017363}, and the notion of spatiotemporal modulations has been theoretically and experimentally utilized to break Lorentz symmetry and instigate nonreciprocal wave transmission. These space-time-periodic systems have ranged in implementation from phononic crystals \cite{Nassar201710,attarzadeh2018non,trainiti2016non} and locally resonant metamaterials \cite{attarzadeh2018wave, attarzadeh2020experimental,nassar2017non}, to stimuli-responsive materials \cite{ansari2017application}. While stiffness is typically the modulation property of choice, given its accessibility and potential experimental tunability, similar studies have looked into the modulation of other inertial properties including mass \cite{huang2019mass, huang2020mass2} and angular momentum in gyroscopic structures \cite{attarzadeh2019gyro}. While all the previous efforts rely on real modulation of material property, the concept of using non-Hermitian modulations has recently been adopted particularly in stiffness-modulated waveguides in the form of imaginary harmonic spatial functions \cite{Riva_non_Her}. Given the fact that such non-Hermitian modulations are invariant under the action of parity and time-reversal operators, the resultant structures are deemed to be $\mathcal{PT}$ symmetric.

$\mathcal{PT}$-symmetric non-Hermitian Hamiltonians, which are capable of possessing real energy spectra, were initially conceived of in quantum physics \cite{bender1998real}. There, a sudden phase transition from the exact to the broken region (or vice versa) was realized by passing a critical threshold in which two eigenvalues and their corresponding eigenvectors coalesce, commonly known as the exceptional point degeneracy \cite{wang2022exceptional}. In the broken phase, eigenvalues emerge as complex conjugate pairs in which modes with positive and negative imaginary components experience exponential amplification and attenuation, respectively. Given the similarity between the Schrödinger equation in quantum mechanics and the wave equation, $\mathcal{PT}$ symmetry was readily extended to optics \cite{ozdemir2019parity,makris2010pt,miri2012bragg,zheng2010non}, acoustics \cite{yang2022design,gu2021controlling,zhu2014p} and elastic systems \cite{geng2021chiral,christensen2016parity,braghini2021non}, primarily through ``gain-loss energy pumping mechanisms" via an odd, imaginary form of modulation. Notable among these are the realizations of unidirectional reflectionless absorbers \cite{feng2013experimental}, microring lasers \cite{ge2011unconventional,hodaei2014parity}, dynamic power oscillations \cite{miroshnichenko2011nonlinearly,ruter2010observation}, nonreciprocal light propagation \cite{chang2014parity}, and invisible acoustic sensors \cite{fleury2015invisible}. In the photonics domain, different approaches have been undertaken to implement $\mathcal{PT}$-symmetric optical systems, including aperiodic temporal modulation \cite{li2020parity}, dynamic gain-loss modulation \cite{song2019direction,liu2020scattering}, and Floquet systems \cite{wang2018photonic,koutserimpas2018parametric}. However, $\mathcal{PT}$-symmetric phononic systems have been significantly less explored and limited to space-periodic modulations. In these systems, an imaginary stiffness modulation of an elastic medium becomes equivalent to a cycle of negative (gain) and positive (loss) damping terms in the motion equations that can be experimentally enacted via negative and positive resistive shunt circuits attached to piezoelectric patches that are bonded to the elastic medium, along with a negative capacitance that amplifies the gain and loss effects as needed \cite{wu2019asymmetric}. Additionally, there remains a large gap between the dynamics, response and behavior of $\mathcal{PT}$-symmetric systems on the one hand, and the wave dispersion patterns that culminate from different modulation forms on the other. To date, the vast majority of efforts have primarily focused on leveraging the features of such systems in the context of scattering properties at the boundaries and downstream of modulated waveguides, with little to no attention to the wave propagation profile inside the medium itself. 

The current work aims to address the aforementioned questions by providing a comprehensive treatment of spatiotemporal stiffness modulations of an elastic medium, ranging from purely real to complex modulations that are used to exploit non-Hermitian degeneracies in $\mathcal{PT}$-symmetric phononic systems. We start by deriving a unifying theory that employs the plane-wave expansion approach to evaluate the dispersion relations of a one-dimensional slender bar, in which the stiffness profile is modulated using a universal, complex spatiotemporal waveform. Developing the analytical framework for the most general case enables us to investigate the dispersion properties for a number of interesting subscenarios and draw comparisons between $\mathcal{PT}$-symmetric systems with different modulation forms, by extracting insight from the evolution of the so-called wave-number gaps, their existence conditions, and their unique behavior associated with the different scenarios. Towards the second half of the paper, we evaluate the effect of the modulation's parameters on both the location and properties of the emergent exceptional points (EPs), by studying and quantifying the directional amplification taking place at such EPs in the presence and absence of temporal modulation of material property. The ensuing analysis interestingly shows that directional amplification of select EPs in time-periodic elastic systems can only be triggered by an input excitation whose frequency signature represents a specific function of the modulating frequency, as will be detailed.

\vspace{-0.3cm}

\section{Theory} 
\label{Sec:DA}

To study elastic wave dispersion in a continuous system with non-Hermitian space-time modulation, we start with the one-dimensional wave equation governing the structural dynamics of an elastic longitudinal bar, given by
\begin{equation}
\label{eq:vib}
    \frac{\partial}{\partial x}  \left[ E(x,t)  \frac{\partial u(x,t)}{\partial x}  \right] 
    = 
    \frac{\partial}{\partial t} \left [\rho(x,t) \frac{\partial u(x,t)}{\partial t} \right ], 
\end{equation}
where $u(x,t)$ describes the bar's displacement at a field point $x$ and a time instant $t$, and is denoted $u$ hereafter for simplicity. $E(x,t)$ and $\rho(x,t)$ represent the space- and time-dependent elastic modulus and mass per unit length, respectively. The non-Hermitian modulations of the bar's elastic modulus and density are given by
\begin{subequations}
\label{eq:modulation}
\textbf{\begin{align} 
    E(x,t) = E_o \left[1+\alpha \cos(\omega_p t - \kappa_p x) + \ii \beta \sin(\omega_p t -\kappa_p x)\right],
    \\
   \rho(x,t) = \rho_o \left[1+\alpha \cos(\omega_p t - \kappa_p x) + \ii \beta \sin(\omega_p t -\kappa_p x)\right],
\end{align}}
\end{subequations}
where $E_o$ and $\rho_o$ represent the average modulus and density values, $\ii=\sqrt{-1}$ is the imaginary unit, $\omega_p$ is the temporal modulation frequency, and $\kappa_p$ is the spatial modulation frequency. The coefficients $\alpha$ and $\beta$ dictate the depths of the real and imaginary spatiotemporal modulations, respectively. 

Owing to the periodicity of $E$ and $\rho$ in both space and time, both can be rewritten via Fourier expansions as:
\begin{subequations}
\label{eq:modulation_exp}
\textbf{\begin{align} 
    E(x,t)=\sum_{r=-\infty}^{\infty} E_r \ee^{\ii r(\omega_pt-\kappa_px)},
    \\
    \rho(x,t)=\sum_{r=-\infty}^{\infty} \rho_r \ee^{\ii r(\omega_pt-\kappa_px)}.
\end{align}}
\end{subequations}
The Fourier coefficients $E_r$ of the elastic modulus are given by
\textbf{\begin{align}
\label{eq:modulation_exp_simp1}
    E_r = \frac{\omega_p}{2\pi} \frac{\kappa_p}{2\pi} &\int_{0}^{\frac{2\pi}{\omega_p}} \int_{0}^{\frac{2\pi}{\kappa_p}} \Re \big(E(x,t)\big) \, \ee^{-\ii r(\omega_pt-\kappa_px)} \, dx \, dt \, 
    \nonumber\\
    &+\ii \frac{\omega_p}{2\pi} \frac{\kappa_p}{2\pi} \int_{0}^{\frac{2\pi}{\omega_p}} \int_{0}^{\frac{2\pi}{\kappa_p}} \Im \big(E(x,t) \big) \, \ee^{-\ii r(\omega_pt-\kappa_px)} \, dx \, dt
\end{align}}
with the operators $\Re ( \cdot )$ and $\Im ( \cdot )$ returning the real and imaginary parts of their arguments, respectively. Substituting Eq.~(\ref{eq:modulation}a) into (\ref{eq:modulation_exp_simp1}) gives
\textbf{\begin{align} 
\label{eq:modulation_exp_simp2}
    E_r =E_o \frac{\omega_p}{2\pi} \frac{\kappa_p}{2\pi} &\int_{0}^{\frac{2\pi}{\omega_p}} \int_{0}^{\frac{2\pi}{\kappa_p}} \big[1+\alpha \cos(\omega_p t - \kappa_p x)\big] \, \ee^{-\ii r(\omega_pt-\kappa_px)} \, dx \, dt \,
    \nonumber\\
    &+\ii E_o \frac{\omega_p}{2\pi} \frac{\kappa_p}{2\pi} \int_{0}^{\frac{2\pi}{\omega_p}} \int_{0}^{\frac{2\pi}{\kappa_p}} \big[\beta \sin(\omega_p t - \kappa_p x)\big] \, \ee^{-\ii r(\omega_pt-\kappa_px)} \, dx \, dt,
\end{align}}
which, following several mathematical manipulations, further simplifies to
\textbf{\begin{align} 
\label{eq:E_final}
    E_r =& E_o \left[\delta_r + \frac{\alpha}{2} (\delta_{r-1} + \delta_{r+1})\, + \frac{\beta}{2} (\delta_{r-1}-\delta_{r+1}) \right],
\end{align}}
where $\delta_{r}$ is Dirac-delta function that is equal to unity for $r=0$ and zero otherwise. With the focus here being on stiffness-modulated structures, we set $\rho(x,t)=\rho_o$ as an invariant parameter and therefore the Fourier coefficient corresponding to the constant density can be defined as $\rho_r = \rho_o \delta_r$. By incorporating the Floquet theorem and implementing the plane-wave expansion (PWE) method \cite{park2021spatiotemporal}, a solution of the following form is realized:
\begin{equation} \label{eq:solutionl}
    u(x,t)=\ee^{\ii (\omega t -\kappa x)}\sum_{s=-\infty}^{\infty} u_s \ee^{\ii s(\omega_pt-\kappa_px)}.
\end{equation}
Here $u_s$ is the amplitude of the $s$ harmonic in the assumed solution. Upon substituting Eqs.~(\ref{eq:modulation_exp}) and (\ref{eq:solutionl}) back into (\ref{eq:vib}), and canceling out the $\ee^{\ii(\omega t -\kappa x)}$ term, the following equality can be obtained:
\begin{align} \label{eq:solutionl_simp1}
    \sum_{r=-\infty}^{\infty}  \sum_{s=-\infty}^{\infty} & \kappa_{[r+s]} \kappa_{[s]} E_r u_s \ee^{\ii (r+s)(\omega_pt-\kappa_px)} = \sum_{r=-\infty}^{\infty} \sum_{s=-\infty}^{\infty} \omega_{(r+s)} \omega_{(s)} \rho_r u_s  \ee^{ \ii (r+s) (\omega_p t-\kappa_p x)}.
\end{align}
Here $\kappa_{[r]}=\kappa + r \kappa_p$ and $\omega_{(r)}=\omega + r \omega_p$ are shorthand notation for wave-number and frequency shifts, respectively, for any integer $r$. To exploit the orthogonality of harmonic functions, both sides of Eq.~(\ref{eq:solutionl_simp1}) are first multiplied by $\ee^{-\ii \ell (\omega_pt-\kappa_px)}$, with $\ell$ being a dummy integer variable, to get 
\begin{align} \label{eq:solutionl_simp2}
    \sum_{r=-\infty}^{\infty}  \sum_{s=-\infty}^{\infty} & \kappa_{[r+s]} \kappa_{[s]} u_s  E_r \ee^{i(r+s-\ell)(\omega_pt-\kappa_px)} = \sum_{r=-\infty}^{\infty} \sum_{s=-\infty}^{\infty} \omega_{(r+s)} \omega_{(s)} u_s \rho_r \ee^{i(r+s-\ell) (\omega_p t-\kappa_p x)},
\end{align}
following which, both sides of Eq.~(\ref{eq:solutionl_simp2}) are averaged over one spatial and temporal period as:
\begin{equation} \label{eq:solutionl_ave}
    \frac{\omega_p}{2\pi}
    \frac{\kappa_p}{2\pi} 
    \int_{0}^{\frac{2\pi}{\omega_p}} 
    \int_{0}^{\frac{2\pi}{\kappa_p}}
    \sum_{r=-\infty}^{\infty}  \sum_{s=-\infty}^{\infty} \big( \kappa_{[r+s]} \kappa_{[s]}   E_r  
    -\omega_{(r+s)} \omega_{(s)} \rho_r \big ) u_s \ee^{i(r+s-\ell) (\omega_p t-\kappa_p x)} dx\, dt = 0.
\end{equation}
Given the orthogonality of harmonic functions, all the summation terms on the left-hand side of Eq.~(\ref{eq:solutionl_ave}) can be zeroed out, except for when $r+s=\ell$ (or equivalently, $r=\ell-s$). As such, Eq.~(\ref{eq:solutionl_ave}) reduces to
\begin{equation}\label{eq:solutionl_orth_simp}
    \sum_{s=-\infty}^{\infty} \big( \kappa_{[r]} \kappa_{[s]} E_{r-s} - \omega_{(r)} \omega_{(s)}\rho_{r-s} \big) u_s =0. 
\end{equation}
Expanding the powers of $\omega$ gives
\begin{align}\label{eq:solutionl_exp}
    \sum_{s=-\infty}^{\infty} \big( \rho_{r-s} \omega^2  + \rho_{r-s} (r+s) \omega_p \omega + \rho_{r-s} rs\omega_p^2 -\kappa_{[+r]} \kappa_{[+s]} E_{r-s} \big) u_s =0. 
\end{align}

The infinite series in Eq.~(\ref{eq:solutionl_exp}) can be truncated by imposing a finite bound $d$ and consequently casting into a familiar matrix eigenvalue problem of the form:
\begin{align} \label{eq_final}
    (\mathbf{A} \, \omega^2 + \mathbf{B}  \, \omega + \mathbf{C})  \, \mathbf{\tilde{u}}=\mathbf{0},
\end{align}
where $\mathbf{\tilde{u}} = \begin{bmatrix} \tilde{u}_{-d}, & \dots, & \tilde{u}_{-1}, & \tilde{u}_{0}, & \tilde{u}_{1}, & \dots, & \tilde{u}_{d} \end{bmatrix}^T$ is the eigenvector, and the entries of matrices $\mathbf{A}$, $\mathbf{B}$ and $\mathbf{C}$ are given by
\begin{subequations}
    \label{eq:coefficients}
\begin{align}
    A_{s,r}&= \mu_{r-s},\\
    B_{s,r}&= (r+s) \omega_p \mu_{r-s},\\
    C_{s,r}&= rs\omega_p^2 \mu_{r-s} - \kappa_{[+r]} \kappa_{[+s]} c_o^2 \gamma_{r-s},
\end{align}
\end{subequations}
where $c_o^2=\frac{E_o}{\rho_o}$ and
\begin{subequations} \label{eq:coefficients_param}
\begin{align}
    \mu_{r-s}&=\frac{\rho_{r-s}}{\rho_{0}}=\delta_{r-s},
    \\
    \gamma_{r-s}&=\frac{E_{r-s}}{E_{0}}=\delta_{r-s} + \frac{\alpha}{2} (\delta_{r-s-1} + \delta_{r-s+1})\, + \frac{\beta}{2} (\delta_{r-s-1}-\delta_{r-s+1}).
\end{align}
\end{subequations}
%
\section{Generalized Modulation Scenarios} 
\label{Sec:Results}

By applying the most general form of harmonic stiffness modulation (real and imaginary variations in both space and time) to derive the eigenvalue problem depicted in Eq.~(\ref{eq_final}), we are able to start from the generalized case and simplify down to specific cases of interest, as will be shown in the following subsections. We start with a couple of simple examples where the modulation is purely real and either space or time periodic to confirm and validate the current framework. Following which, we proceed to uncharted scenarios that will reveal the intriguing physics of complexly modulated elastic systems ranging from symmetry to existence conditions of different types of dispersion gaps. In the following, $E_o = \rho_o = 1$ and all the parameters are expressed in arbitrary units.

\subsection{Cases 1 and 2: $E(x,t)=E_o \left[1+\alpha \cos(-\kappa_p x)\right]$ and $E(x,t)=E_o \left[1+\alpha \cos(\omega_p t)\right]$, respectively}

\begin{figure}[h!]
\centering
\includegraphics[width=\textwidth]{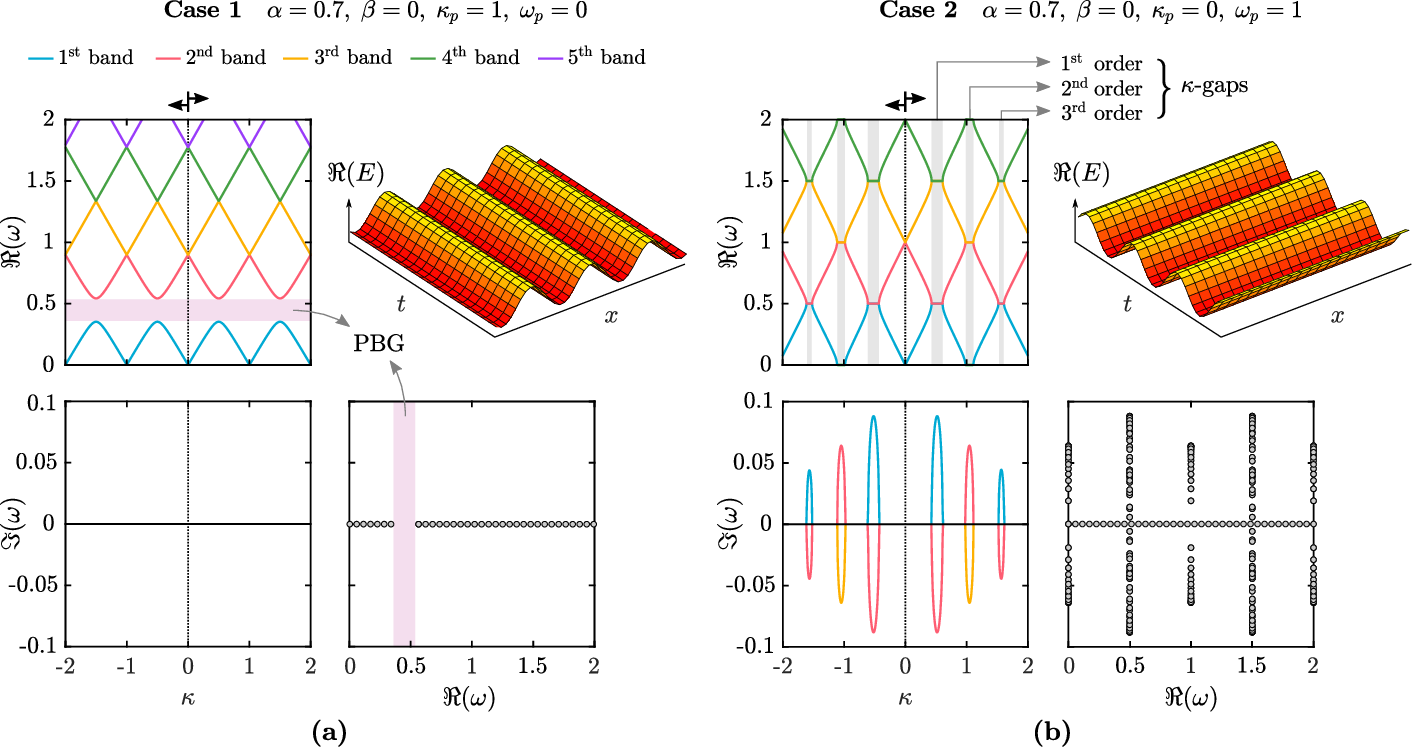}
\caption{Dispersion analysis for the systems described by (a) case 1 in which $\alpha=0.7$, $\beta=0$, $\kappa_p=1$, and $\omega_p=0$; and (b) case 2 in which $\alpha=0.7$, $\beta=0$, $\kappa_p=0$, and $\omega_p=1$. The top left plot depicts the band structure, i.e., the real frequency $\Re(\omega)$ versus the wave number $\kappa$ for each case. The bottom left and bottom right plots provide the variation of the imaginary frequency $\Im(\omega)$ versus $\kappa$ and $\Re(\omega)$, respectively, for each case. A graphical representation of the real elastic modulus modulation $\Re(E)$ in space and time is provided in the top right corner of each case ($E_o = \rho_o = 1$).}
\label{fig:case1&2}
\end{figure}

\hspace{0.9cm} The stiffness modulation profile represented by case 1 describes a traditional phononic bar, albeit with a harmonically varying modulus rather than discrete periodic impedance mismatches. The emergence and behavior of Bragg band gaps in infinite \cite{khelif2006complete, hsu2006efficient, laude2009evanescent} and finite \cite{al2017pole, Nouh2017, bastawrous2022closed} phononic systems have been thoroughly studied over the past two decades. The approach adopted here is a free-wave approach where a real wave number $\kappa$ is fed into the dispersion relation to obtain the real and imaginary components of the output frequency $\omega$. The left column of Fig.~\ref{fig:case1&2}a pits the real and imaginary components of the eigenfrequency, $\Re(\omega)$ and $\Im(\omega)$, against the real wave number $\kappa$ for case 1 with $\alpha=0.7$, $\beta=0$, $\kappa_p=1$, and $\omega_p=0$. The former shows the first five dispersion bands (in a color coded order) and reveals the existence of a phononic band gap (PBG) between the first and second bands, while the latter shows a zero imaginary component of frequency across the entire wave-number space, confirming the expected lack of temporal attenuation-amplification in the nondissipative time-invariant elastic medium. 
Most recently, the emergence of $\kappa$ gaps, a region in the wave-number space where two bands coalesce with complex conjugate eigenfrequencies, has been reported in elastic structures with complex spatially modulated stiffness profiles \cite{Riva_non_Her}, as well as spatially uniform elastic structures with real time-modulated stiffness profiles \cite{trainiti2019time}.

Figure~\ref{fig:case1&2}b shows case 2 with $\alpha=0.7$, $\beta=0$, $\kappa_p=0$, and $\omega_p=1$. By inspecting Eqs.~(\ref{eq_final}) through (\ref{eq:coefficients_param}), it can be deduced that the $\mathbf{A}$, $\mathbf{B}$ and $\mathbf{C}$ matrices are both real and symmetric in case 2. Because of the properties of quadratic eigenvalue problems, eigenvalues of such a system can either be real values or complex conjugates \cite{tisseur2001quadratic}. As a result, we witness the formation of several $\kappa$ gaps as shown in the top left panel of Fig.~\ref{fig:case1&2}b. These $\kappa$ gaps can be classified as $\kappa$ gaps of the $n^{\text{th}}$ order, and sorted based on their proximity to the $\kappa=0$ axis. Henceforth, we refer to the $\kappa$ gaps closest to the $\kappa=0$ axis as first-order gaps. These first-order gaps form at the interface between the first and second, third and fourth, fifth and sixth bands, etc. We refer to following set of $\kappa$ gaps as second-order gaps. These are the first set of gaps that form at the interface between the second and third, fourth and fifth, fifth and sixth bands, etc, and so forth. The formation of $\kappa$ gaps corresponds to the onset of temporal amplification (or attenuation) regions captured by the positive (or negative) imaginary component of the eigenfrequency, $\Im(\omega)$ (bottom left panel of Fig.~\ref{fig:case1&2}b). It can be seen that both the width (with respect to the wave-number axis) and the amplification-attenuation level, i.e., $|\Im(\omega)|$, of the $\kappa$ gaps decrease as they move further away from $\kappa=0$ and their order increases. It can also be seen that $\kappa$ gaps of the same order that span the same wave-number range but occur at different frequencies, have identical amplification-attenuation levels. As a result, $\Im(\omega)$ corresponding to the first-order $\kappa$ gap at the interface between the first and second dispersion bands lies on top of the first-order $\kappa$ gap between the third and fourth bands.

The bottom right panel of Fig.~\ref{fig:case1&2}b plots the imaginary component of each eigenfrequency as a function of its real component, and confirms the presence of several complex conjugate eigenfrequencies that share the same real value but different imaginary parts at each $\kappa$ gap. The figure also reveals that even-ordered $\kappa$ gaps correspond to $\Re(\omega)=\vartheta/2$ for $\vartheta \in \{0, 2, 4, \dots \}$, while odd-ordered $\kappa$ gaps correspond to $\Re(\omega)=\varphi/2$ for $\varphi \in \{1, 3, 5, \dots \}$. Finally, the frequency of each $\kappa$ gap is identical for left- and right-going waves, and the symmetry of the $\Re(\omega)$-$\kappa$ plot about the $\kappa=0$ axis is preserved.

\vspace{-0.25cm}

\subsection{Cases 3 and 4: $E(x,t)=E_o \left[1+\alpha \cos(\omega_p t - \kappa_p x)\right]$ for $\nu<1$ and $\nu>1$, respectively}

\vspace{-0.25cm}

\begin{figure}[h!]
\centering
\includegraphics[width=\textwidth]{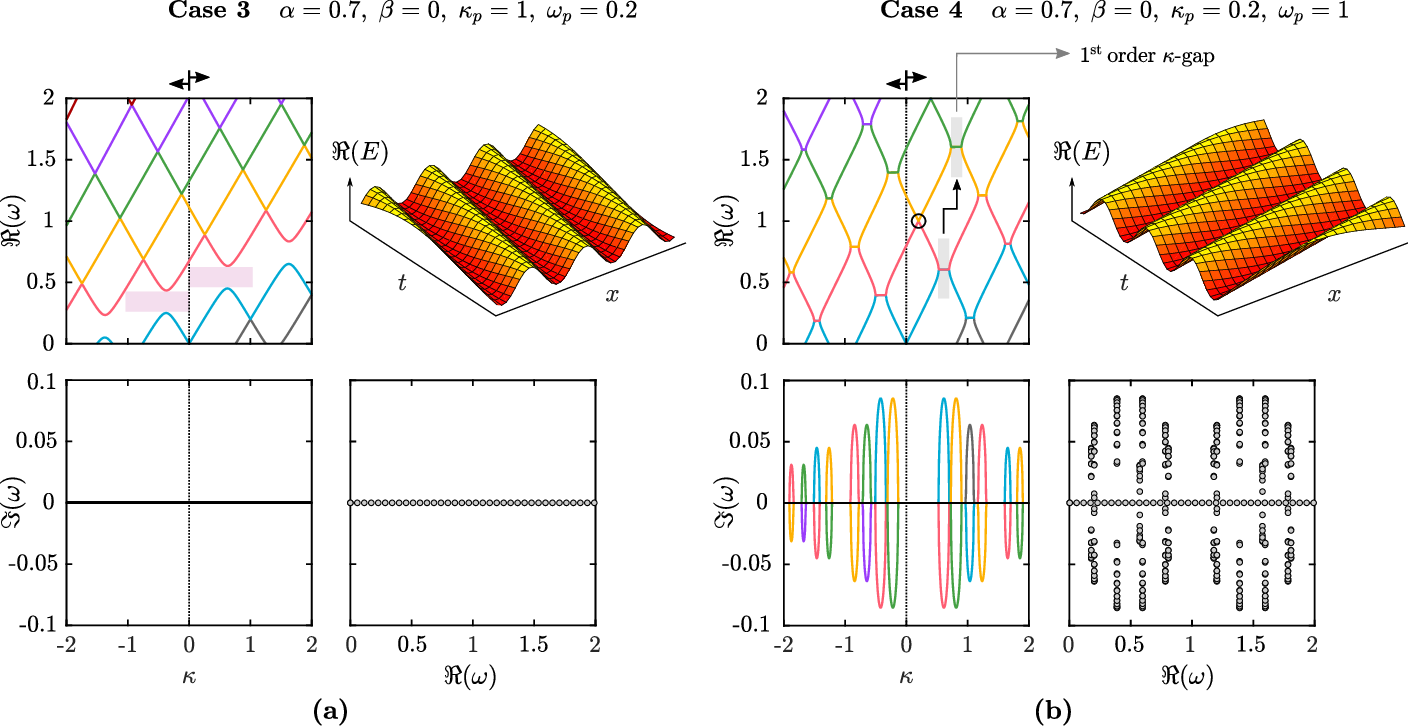}
\caption{Dispersion analysis for the systems described by (a) case 3, in which $\alpha=0.7$, $\beta=0$, $\kappa_p=1$, and $\omega_p=0.2$; and (b) case 4, in which $\alpha=0.7$, $\beta=0$, $\kappa_p=0.2$, and $\omega_p=1$. The top left plot depicts the band structure, i.e., the real frequency $\Re(\omega)$ versus the wave number $\kappa$ for each case. The bottom left and bottom right plots provide the variation of the imaginary frequency $\Im(\omega)$ versus $\kappa$ and $\Re(\omega)$, respectively, for each case. A graphical representation of the real elastic modulus modulation $\Re(E)$ in space and time is provided in the top right corner of each case ($E_o = \rho_o = 1$).}
\label{fig:case3&4}
\end{figure}

\hspace{0.4cm} For an elastic bar with a spatiotemporally modulated stiffness, we define the modulation speed $v_p$ as the ratio between the temporal modulation frequency $\omega_p$ and the spatial modulation frequency $\kappa_p$, and the modulation speed ratio as $\nu = v_p / c_o$, where $c_o$ denotes the sonic speed in the medium. Figure~\ref{fig:case3&4}a shows a phononic bar with a purely real space-time-periodic elastic modulus, i.e., case 3,  with $\alpha=0.7$, $\beta=0$, $\kappa_p=1$, and $\omega_p=0.2$. The previous parameters amount to a subsonic modulation speed ($\nu < 1$) that gives rise to a direction-dependent PBG, i.e., a band gap that spans different frequencies for positive and negative values of $\kappa$ \cite{trainiti2016non}, a behavior which has recently been exploited to realize nonreciprocal wave transmission or a diodelike behavior in elastic metamaterials \cite{attarzadeh2020experimental}. As the value of $\nu$ increases, the modulation speed approaches that of the wave. The $\nu=1$ condition denotes what is referred to as a \textit{luminal} modulation in photonics \cite{galiffi2019broadband}. Further details about the implications of this condition can be found in the work of Cassedy and Oliner \cite{cassedy1962temporal} and Cassedy \cite{cassedy1967dispersion}. The dispersive behavior of the medium drastically changes beyond that point, once $\nu$ exceeds unity, as demonstrated by case 4.

In case 4, the parameters $\alpha=0.7$, $\beta=0$, $\kappa_p=0.2$, and $\omega_p=1$ lead to a supersonic modulation speed ($\nu > 1$). As a result, $\kappa$ gaps that are asymmetric with respect to the $\kappa=0$ axis are generated as can be seen in Fig.~\ref{fig:case3&4}b. These $\kappa$ gaps are not only asymmetric with respect to the wave-number axis, but also the frequencies they correspond to differ for left- and right-going waves, indicating a nonreciprocal wave amplification-attenuation capability. It is important to point out that the symmetry breakage of the band structure, and the overall tilting associated with it, slightly alters the location of $\kappa$ gaps of the same order that now no longer span the same range of $\kappa$ values. For visualization, this is graphically pointed out in the horizontal shift between the lower-frequency and the higher-frequency first-order $\kappa$ gaps in the $\Re(\omega)$-$\kappa$ plot of Fig.~\ref{fig:case3&4}b. As a result, while these two $\kappa$ gaps no longer share the same proximity to the $\kappa=0$ axis, we still define both of them as first-order $\kappa$ gaps since they are the closest gaps to the $\kappa=0$ axis that take place at the interface between the first and second bands, and the third and fourth, respectively. (The same definition extends to the $\kappa$ gaps observed in the band structures of cases 7 and 8 that also exhibit tilting.) The circled region in the same figure denotes the shifted location of the meeting point between the second and third bands, which has deviated from the $\kappa=0$ axis. Finally, similar to case 2, we note that both the width (with respect to wave-number axis) and the amplification-attenuation level, i.e., $|\Im(\omega)|$, of the $\kappa$ gaps become smaller for higher-order gaps. In other words, they gradually shrink and eventually vanish as we move further away from the $\kappa=0$ axis.

\subsection{Cases 5 and 6: $E(x,t)=E_o \left[1 + \ii \beta \sin( - \kappa_p x)\right]$ and $E(x,t)=E_o \left[1 + \ii \beta \sin(\omega_p t)\right]$}, respectively

\begin{figure}[h!]
\centering
\includegraphics[width=\textwidth]{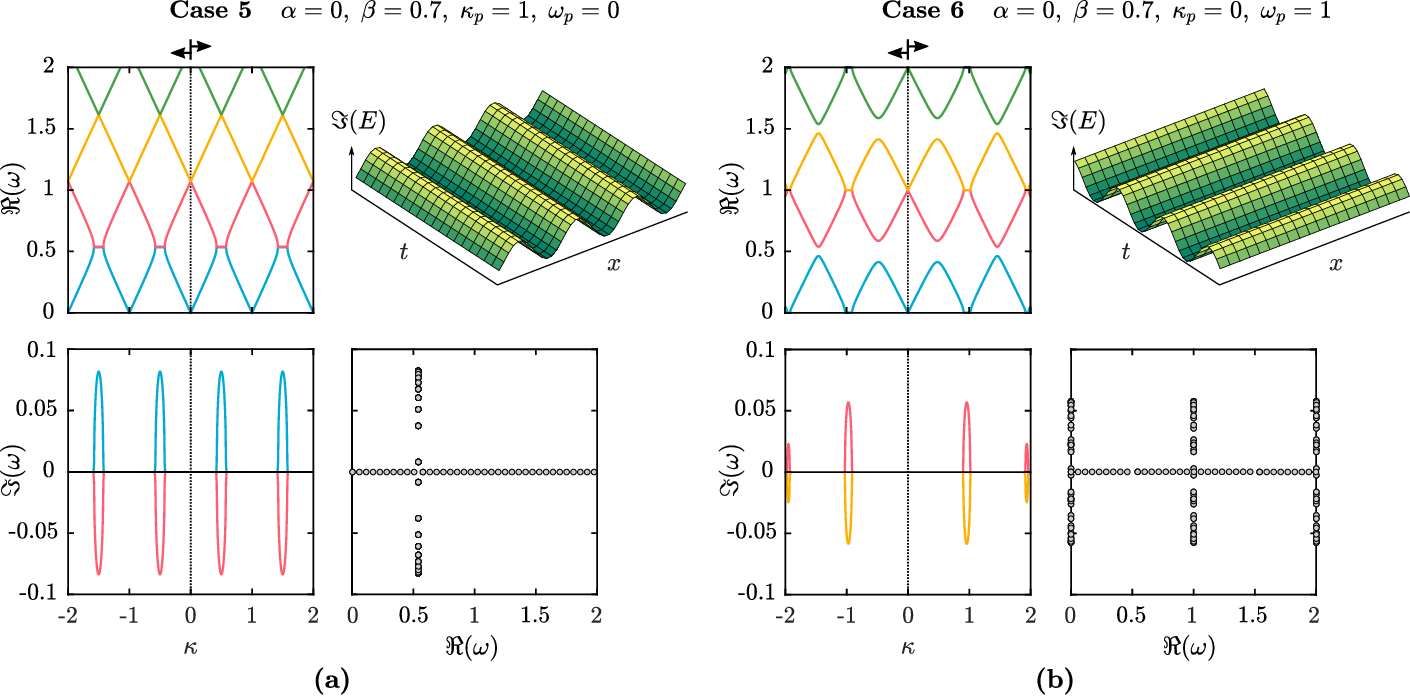}
\caption{Dispersion analysis for the systems described by (a) case 5, in which $\alpha=0$, $\beta=0.7$, $\kappa_p=1$, and $\omega_p=0$; and (b) case 6, in which $\alpha=0$, $\beta=0.7$, $\kappa_p=0$, and $\omega_p=1$. The top left plot depicts the band structure, i.e., the real frequency $\Re(\omega)$ versus the wave number $\kappa$ for each case. The bottom left and bottom right plots provide the variation of the imaginary frequency $\Im(\omega)$ versus $\kappa$ and $\Re(\omega)$, respectively, for each case. A graphical representation of the imaginary elastic modulus modulation $\Im(E)$ in space and time is provided in the top right corner of each case ($E_o = \rho_o = 1$).}
\label{fig:case5&6}
\end{figure}

\hspace{0.8cm} In cases 5 and 6, we examine a phononic bar with a stiffness modulation profile that includes a purely imaginary spatial modulation term and a purely imaginary temporal modulation term, respectively. In both of these cases, we utilize an odd function, namely $\sin(\cdot)$, to synthesize a gain-loss mechanism and trigger periodic energy pumping into and out of the system, akin to non-Hermitian modulations of optical systems \cite{song2019direction}. A close inspection of Eq.~(\ref{eq_final}) for case 5 reveals a suppression of the matrix $\mathbf{B}$ due the fact that $\omega_p = 0$. Consequently, Eq.~(\ref{eq_final}) becomes a linear eigenvalue problem. Furthermore, it can be observed that while both $\mathbf{A}$ and $\mathbf{C}$ are real, only $\mathbf{A}$ is symmetric. The role of $\beta$ in breaking the symmetry of $\mathbf{C}$ becomes evident, as it converts a conventional Hermitian system to a non-Hermitian (cyclic) system, in which $\mathbf{C}$ is a summation of Hermitian and skew-Hermitian matrices. Figure~\ref{fig:case5&6}a depicts case 5 with $\alpha=0$, $\beta=0.7$, $\kappa_p=1$, and $\omega_p=0$. Despite the presence of a spatial modulation, we note the absence of any PBGs when such modulation is purely imaginary. And despite the absence of a temporal modulation, the structure exhibits some $\kappa$ gaps. In this scenario, four observations can be made about the emerging $\kappa$ gaps, all of which are notably distinct from the behavior of $\kappa$ gaps obtained via real temporal (case 2) or real spatiotemporal (case 4) modulations.

\begin{enumerate}
    \item The $\Re(\omega)$-$\kappa$ plot shows that $\kappa$ gaps only form at a single frequency ($\omega=0.5$).
    \item The $\Re(\omega)$-$\kappa$ plot also shows that all even-ordered $\kappa$ gaps cease to exist.
    \item The $\Im(\omega)$-$\kappa$ plot shows that peak amplification-attenuation levels associated with the $\kappa$ gaps, i.e., $\pm |\Im(\omega)|$, do not increase for higher-order gaps as we move further away from the $\kappa=0$ axis, and instead remain constant.
    \item Both the $\Re(\omega)$-$\kappa$ and $\Im(\omega)$-$\kappa$ plots show that the width of the $\kappa$ gaps remains constant as we move further away from the $\kappa=0$ axis.
\end{enumerate}

\begin{figure}[h!]
\centering
\includegraphics[width=\linewidth]{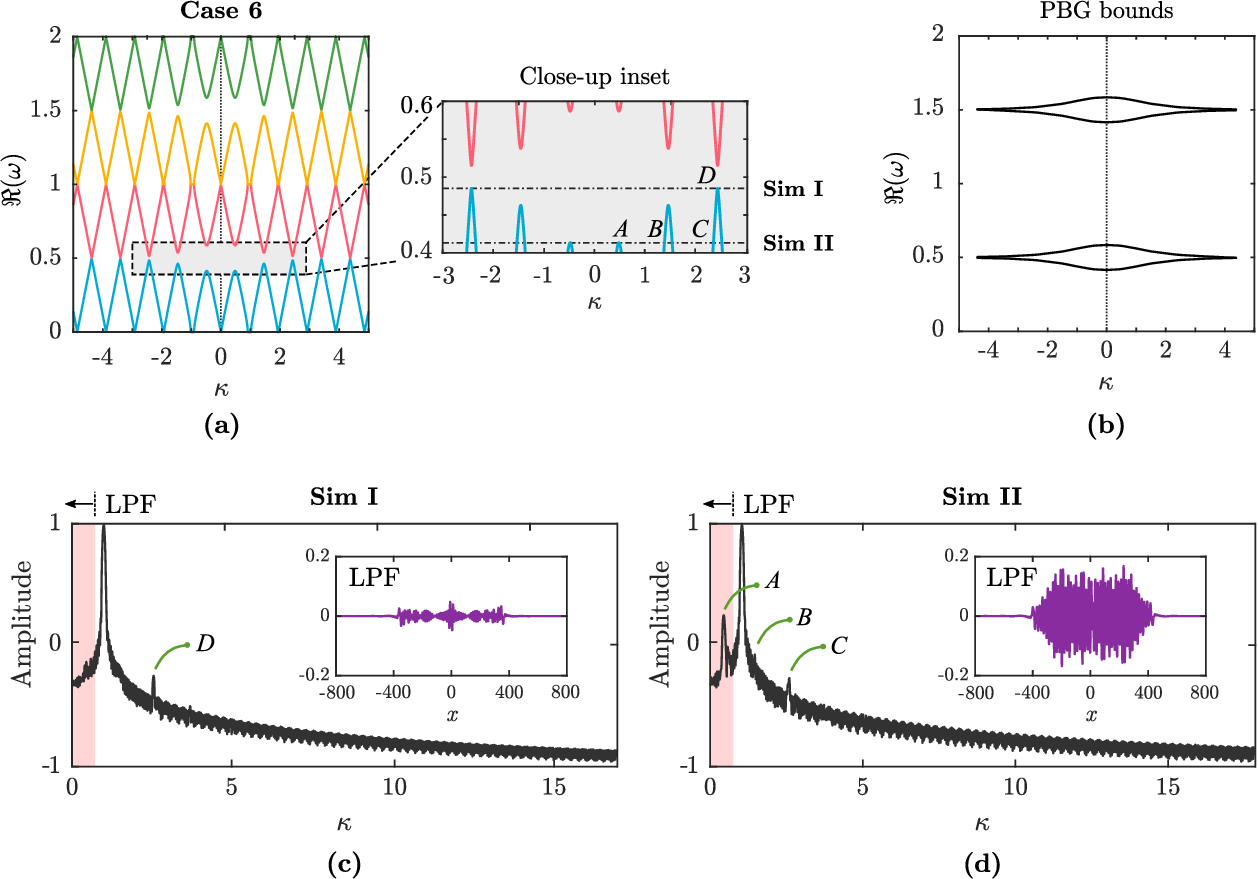}
\caption{(a) Dispersion diagram of case 6 with an extended wave-number axis. Close-up inset shows the gradually narrowing band gap region and indicates the excitation frequencies for two simulations labeled sim I and sim II. (b) Bounds of the two PBGs forming between the first and second bands and between the third and forth bands. PBGs shrink at increasing distance from the $\kappa = 0$ axis and eventually close. (c),(d) Wave-number spectrum (i.e., spatial FFT) of the response of a finite-sized bar made of 50 unit cells to a midpoint excitation at frequencies of (c) $\Re(\omega)=0.48$ (sim I) and (d) $\Re(\omega)=0.42$ (sim II). Markers at \textit{A}, \textit{B}, \textit{C}, and \textit{D} confirm peaks corresponding to the dispersion crossings shown in the close-up inset of (a). Insets in (c) and (d) display snapshots of the spatial wave propagation in the finite bar following a low-pass filter (LPF) admitting the $\kappa < 0.75$ portion, or the shaded region, of the wave-number spectrum.}
\label{fig:PBG_shrink}
\end{figure}

Figure~\ref{fig:case5&6}b represents case 6 with $\alpha=0$, $\beta=0.7$, $\kappa_p=0$, and $\omega_p=1$. Here, a simultaneous combination of PBGs and even-ordered $\kappa$ gaps takes place. Additionally, it is observed that the PBGs and $\kappa$ gaps take turns and are alternating in nature, with only one of the two happening between each two consecutive dispersion bands. The bottom right panel of Fig.~\ref{fig:case5&6}b confirms that the $\kappa$ gaps correspond to $\Re(\omega)=\vartheta/2$ for $\vartheta \in \{0, 2, 4, \dots \}$, and the bottom left panel shows that contrary to Fig.~\ref{fig:case5&6}a, their peak amplification-attenuation levels decrease as we move further away from the $\kappa=0$ axis towards higher-order $\kappa$ gaps. The emergence of PBGs in case 6 is especially noteworthy given the lack of a spatial modulus variation that is typically a hallmark feature of structures with Bragg band gaps. Interestingly, the behavior of these PBGs mimics that of the $\kappa$ gaps in case 2, where the same temporal modulation of the elastic modulus was imposed as a real term. Namely, the frequency range of the PBGs shrinks as we move further from the $\kappa=0$ axis until the gap eventually closes. Figure~\ref{fig:PBG_shrink}a provides the dispersion diagram of case 6 with an extended wave-number axis range, and the adjacent close-up inset confirms the gradual narrowing of the space between the two dispersion branches flanking the PBG as $|\kappa|$ increases. PBG bounds obtained by tracing the edges of these two branches are shown in Fig.~\ref{fig:PBG_shrink}b. To illustrate the manifestation of such PBGs that vary with the wave number, a finite bar consisting of 50 unit cells (i.e., modulation cycles) is excited at its midpoint via two different frequencies, namely $\Re(\omega)=0.48$ (sim I) and $0.42$ (sim II). The two simulations are also indicated in the close-up inset of Fig.~\ref{fig:PBG_shrink}a where it can be seen that sim I lies within the space of the PBG for low values of $\kappa$ and does not hit the lowest dispersion branch until $\kappa = 2.5$, marked as point \textit{D}. On the other hand, owing to its excitation frequency, sim II is unable to evade the same dispersion branch and intersects with it at the locations marked \textit{A}, \textit{B}, and \textit{C}. The full wavenumber spectrum of both simulations is shown in Figs.~\ref{fig:PBG_shrink}c and d, and shows the implication of this behavior on the finite structural response. The highest peak in both plots takes place at the same $\kappa$ value that corresponds to (and is caused by) the modulation frequency of $\omega_p = 1$. Aside from this peak, the bar's response in sim I in Fig.~\ref{fig:PBG_shrink}c is insignificant albeit for the $\kappa$ value at point \textit{D}, whereas contributions from wave numbers at locations \textit{A}, \textit{B}, and \textit{C} can be clearly seen in the bar's response in sim II, shown in Fig.~\ref{fig:PBG_shrink}d; both observations being very consistent with Figs.~\ref{fig:PBG_shrink}a. Contributions from higher wave numbers (i.e., $\kappa>3$) are found to be negligible in both cases. As a final confirmation, the insets in Figs.~\ref{fig:PBG_shrink}c and d show a single snapshot of the spatial wave propagation of both excitations in the finite bar following a low-pass filter only admitting the $\kappa < 0.75$ portion of the wave-number spectrum. The sim II inset in Fig.~\ref{fig:PBG_shrink}d shows the unimpeded propagation of the wave in the finite medium consistent with the lack of a PBG at location \textit{A}, while that in Fig.~\ref{fig:PBG_shrink}c shows an insignificant response since the excitation frequency of sim I squarely falls inside the PBG in the $\kappa < 0.75$ range.

\vspace{0.3cm}

\subsection{Cases 7 and 8: $E(x,t)=E_o \left[1 + \ii \beta \sin(\omega_p t - \kappa_p x)\right]$ for $\nu<1$ and $\nu>1$, respectively}

\hspace{0.4cm} The last two scenarios combine the imaginary spatial and temporal modulations investigated in cases 5 and 6 in one term to capture the behavior of a phononic bar with a purely imaginary space-time-periodic elastic modulus profile. Analogous to cases 3 and 4, we investigate the behavior of this structure at subsonic ($\nu<1$, case 7) and supersonic ($\nu>1$, case 8) modulation speeds. The parameters for case 7 are: $\alpha=0$, $\beta=0.7$, $\kappa_p=1$, and $\omega_p=0.2$. The system in this scenario shares select features from cases 4 (real space-time periodic) and 5 (imaginary space periodic) and represents a hybrid combination thereof. Similar to case 4, case 7 exhibits asymmetric $\kappa$ gaps, both with respect to the $\kappa=0$ axis and with respect to the frequency $\Re(\omega)$ at which they occur for left- and right-going waves. However, similar to case 5, we observe: (1) a lack of even-ordered $\kappa$ gaps, (2) a constant level of amplification-attenuation, $\Im(\omega)$, for all $\kappa$ gaps, and (3) a constant width or $\kappa$ range spanned by all the $\kappa$ gaps regardless of order.

\begin{figure}[h!]
\centering
\includegraphics[width=\textwidth]{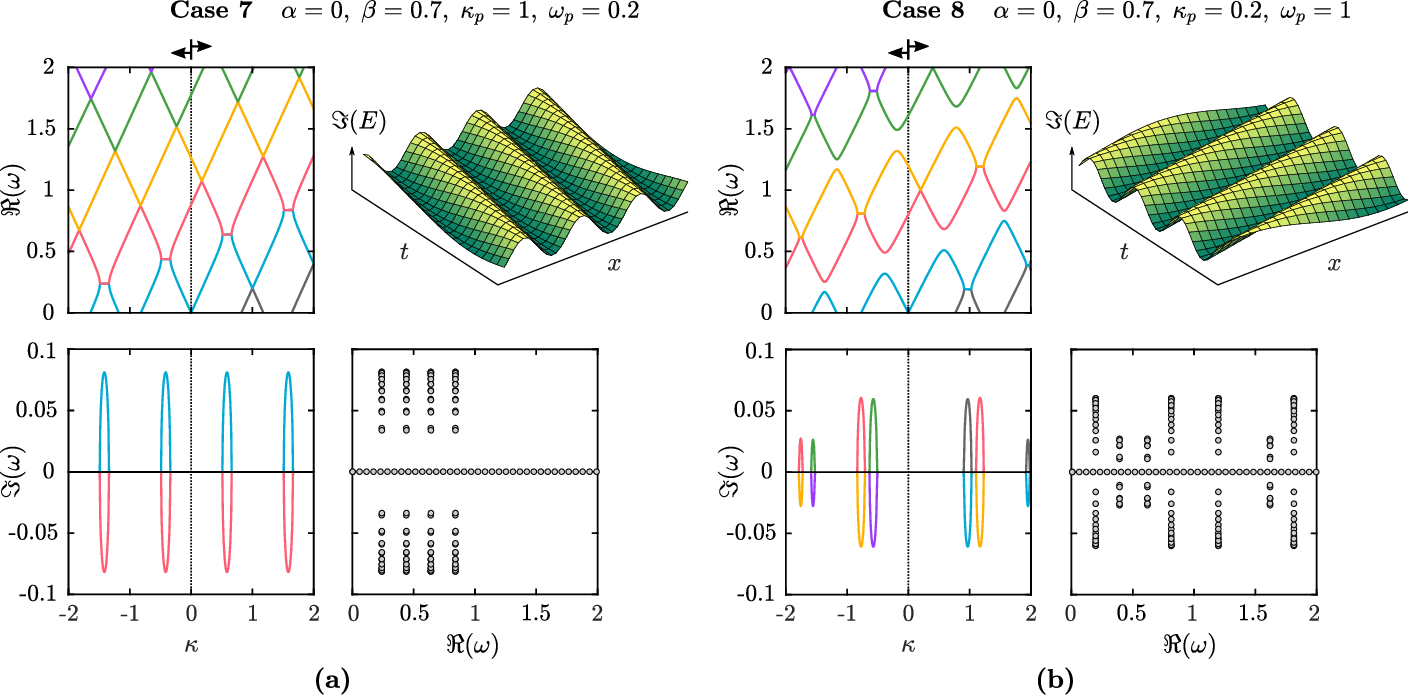}
\caption{Dispersion analysis for the systems described by (a) case 7, in which $\alpha=0$, $\beta=0.7$, $\kappa_p=1$, and $\omega_p=0.2$; and (b) case 8, in which $\alpha=0$, $\beta=0.7$, $\kappa_p=0.2$, and $\omega_p=1$. The top left plot depicts the band structure, i.e., the real frequency $\Re(\omega)$ versus the wave number $\kappa$ for each case. The bottom left and bottom right plots provide the variation of the imaginary frequency $\Im(\omega)$ versus $\kappa$ and $\Re(\omega)$, respectively, for each case. A graphical representation of the imaginary elastic modulus modulation $\Im(E)$ in space and time is provided in the top right corner of each case ($E_o = \rho_o = 1$).}
\label{fig:case7&8}
\end{figure}

Figure~\ref{fig:case7&8}b provides the behavior of case 8 with $\alpha=0$, $\beta=0.7$, $\kappa_p=0.2$, and $\omega_p=1$, which represents a supersonic modulation ($\nu>1$) of the imaginary space-time-periodic stiffness profile. The system in this case largely behaves in a manner similar to case 6, but with a nonreciprocal tilt of the overall band structure as expected. Specifically, we observe the absence of all odd-ordered $\kappa$ gaps as well as an alternating cycle of nonreciprocal PBGs and even-ordered $\kappa$ gaps as we move up the real frequency axis. The width of the even-ordered $\kappa$ gaps narrows for higher-order gaps. Additionally, also similar to case 6, the frequency range of the PBGs also shrinks as we move further from the $\kappa=0$ axis until the gap eventually closes. 

As a final note, it should be pointed out that since eigenfrequencies inside $\kappa$ gaps are represented by complex conjugates, the amplification and attenuation levels for a specific $\kappa$ gap will remain identical for all and any cases. 

\subsection{Verification of $\kappa$ gaps and recovery of the imaginary eigenfrequency component of a finite stiffness-modulated system}

\begin{figure}[h!]
\centering
\includegraphics[width=1\linewidth]{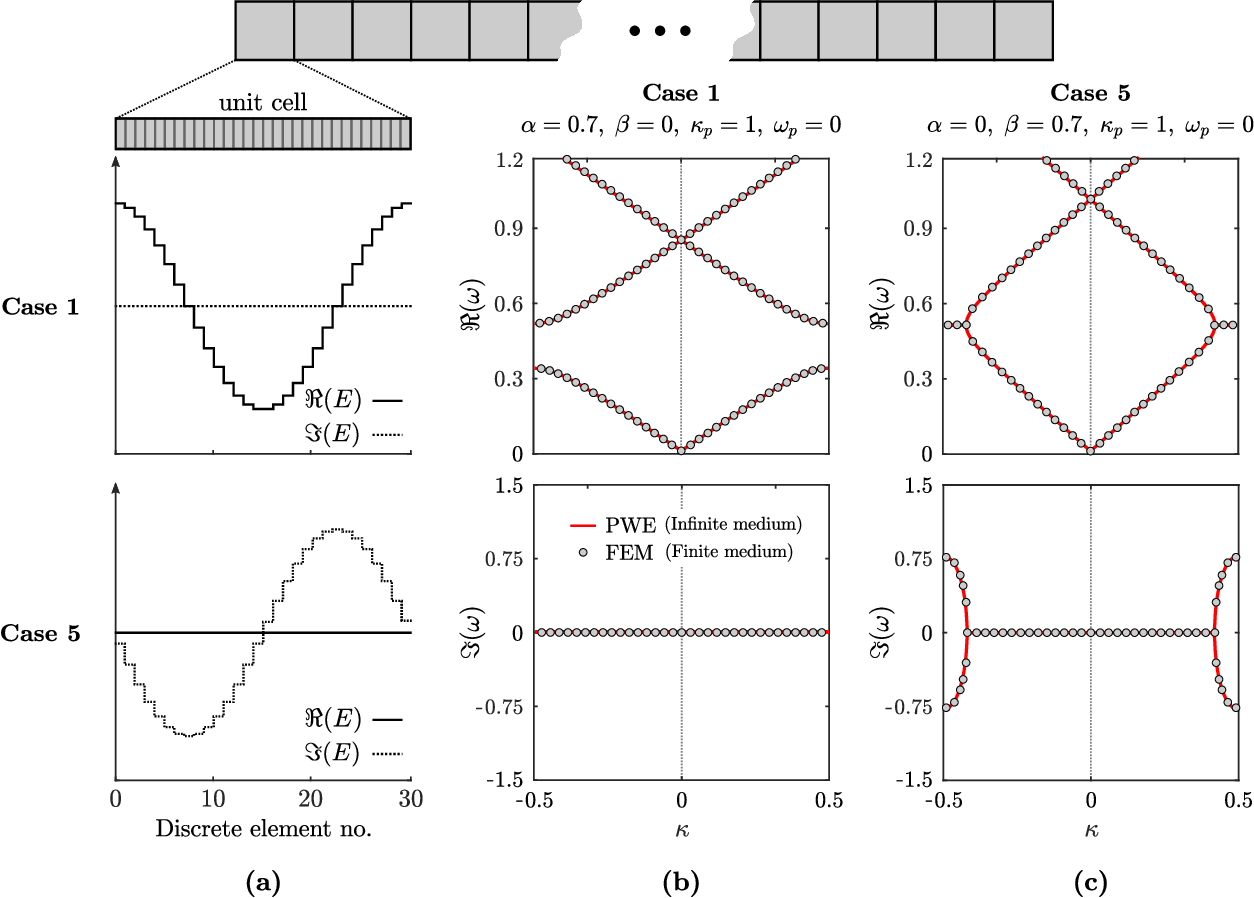}
\caption{Discrete dispersion data obtained from finite realizations of case 1, in which $\alpha=0.7$, $\beta=0$, $\kappa_p=1$, and $\omega_p=0$; and case 5, in which $\alpha=0$, $\beta=0.7$, $\kappa_p=1$, and $\omega_p=0$. (a) Discretized elastic modulus profile over a single spatial modulation cycle spanning a unit cell of 30 finite elements for case 1 (top) and case 5 (bottom). (b),(c) The $\Re(\omega)$-$\kappa$ and $\Im(\omega)$-$\kappa$ diagrams for case 1 (left) and case 5 (right). Finite structure eigenfrequencies (via FEM) are denoted by circles and the infinite band structure (via PWE) is denoted by solid lines.}
\label{fig:finite}
\end{figure}

The formation of $\kappa$ gaps and the accompanying amplification-attenuation regions in the $\Im(\omega)$-$\kappa$ data presented so far have been predicted using an infinite system approach, which adopts the Floquet theorem augmented with a PWE method, as outlined in Section~\ref{Sec:DA}. As a result, it is incumbent upon us to examine and verify the onset of such amplification-attenuation regimes in finite realizations of such stiffness-modulated systems. For brevity, we focus here on cases 1 and 5 that represent real and imaginary spatial modulations of the time-invariant elastic modulus of a one-dimensional phononic bar. To verify the imaginary component of the eigenfrequencies, $\Im(\omega)$, using the finite element method (FEM), we consider a finite bar of 50 unit cells where each cell spans one full cycle of the spatial modulation of the elastic modulus $E$ and is discretized using 30 finite elements, as depicted in Fig.~\ref{fig:finite}a. As a result, the finite structure contains a total of $50 \times 30 = 1,500$ finite elements that are modeled using conventional two-noded 1D bar elements and are assembled to form the equation of motion of the entire structure. The eigenfrequencies are obtained from the free vibration eigenvalue problem $[\mathbf{M}^{-1} \mathbf{K}] \mathbf{u} = \omega^2 \mathbf{u}$, where $\mathbf{u}$ is the discretized displacement field vector, and $\mathbf{M}$ and $\mathbf{K}$ are the overall mass and stiffness matrices, respectively. The resultant eigenfrequencies are split into $\Re(\omega)$ and $\Im(\omega)$ that are plotted in the top and bottom panels of Figs.~\ref{fig:finite}b and c, representing close-up regions of Figs.~\ref{fig:case1&2}a and \ref{fig:case5&6}a. It should be noted that the value of $\kappa$ corresponding to each eigenfrequency is computed from a spatial Fourier transform of the corresponding displacement field of the 1500-element structure. Excellent agreement can be seen between the infinite dispersion bands obtained for both cases and the discrete frequency-wave number obtained from the finite structures representing both cases, including the $\kappa$ gap regions of case 5. More importantly, the bottom panels of Figs.~\ref{fig:finite}b and c, verify the amplification-attenuation levels, i.e., $\Im(\omega)$, corresponding to the $\kappa$ gaps of case 5, in addition to confirming the lack of such gaps in case 1 throughout the entire $\kappa$ space.

\vspace{-0.5cm}

\section{Conditions for Directional Amplification} 

Beyond the eight distinct modulation categories represented by cases 1 through 8, any other form of stiffness modulation can be written as a combination of these cases, and the behavior resulting therefrom will generally depend on two primary factors: the first is whether the absolute value of the ratio $\beta/\alpha$ is less than, greater than, or equal to $1$, and the second is whether the modulation speed ratio $\nu$ is less than, greater than, or equal to $1$. Let us consider sufficiently weak interactions, which enable first-order harmonics to make accurate representations of the system at hand. As a result, we truncate the infinite matrices of Eq.~(\ref{eq:coefficients}) by setting $d=1$. This leads to a homogeneous set of equations governing the free vibrations of a dynamic system with three coupled harmonic oscillators, that are described by
\begin{align} \label{eq:finite}
    (\mathbf{A}_{\lfloor d=1 \rfloor} \, \omega^2 + \mathbf{B}_{\lfloor d=1 \rfloor}  \, \omega + \mathbf{C}_{\lfloor d=1 \rfloor})  \, \mathbf{\tilde{u}}_{\lfloor d=1 \rfloor} = \mathbf{0},
\end{align}
where $\mathbf{\tilde{u}}_{\lfloor d=1 \rfloor} = \begin{bmatrix}
    \tilde{u}_{-1} & \tilde{u}_{0} &  \tilde{u}_{1}
    \end{bmatrix}^T$, and

\begin{subequations}
\label{eq:finite2}
\begin{align}
 \mathbf{A}_{\lfloor d=1 \rfloor} = 
    \begin{bmatrix}
    1 & 0 & 0\\
    0 & 1 & 0\\
    0 & 0 & 1\\
    \end{bmatrix},
\end{align}
\begin{align}
\mathbf{B}_{\lfloor d=1 \rfloor} = 
 \begin{bmatrix}
    -2 \omega_p & 0 & 0\\
    0 & 0 & 0\\
    0 & 0 & 2\omega_p\\
    \end{bmatrix},
\end{align}
\begin{align}
    \mathbf{C}_{\lfloor d=1 \rfloor} =
    \begin{bmatrix}
    \omega_p^2 - c_o^2 (\kappa-\kappa_p)^2 & -c_o^2 \kappa (\kappa-\kappa_p) \frac{\alpha+\beta}{2} & 0\\
    -c_o^2 \kappa (\kappa-\kappa_p)  \frac{\alpha-\beta}{2} & -c_o^2 \kappa^2 & -c_o^2 \kappa (\kappa+\kappa_p) \frac{\alpha+\beta}{2}\\
    0 & -c_o^2 \kappa (\kappa+\kappa_p) \frac{\alpha-\beta}{2} &  \omega_p^2 - c_o^2 (\kappa+\kappa_p)^2 \\
    \end{bmatrix}
\end{align}
\end{subequations}
\begin{figure}[h!]
\centering
\includegraphics[width=\linewidth]{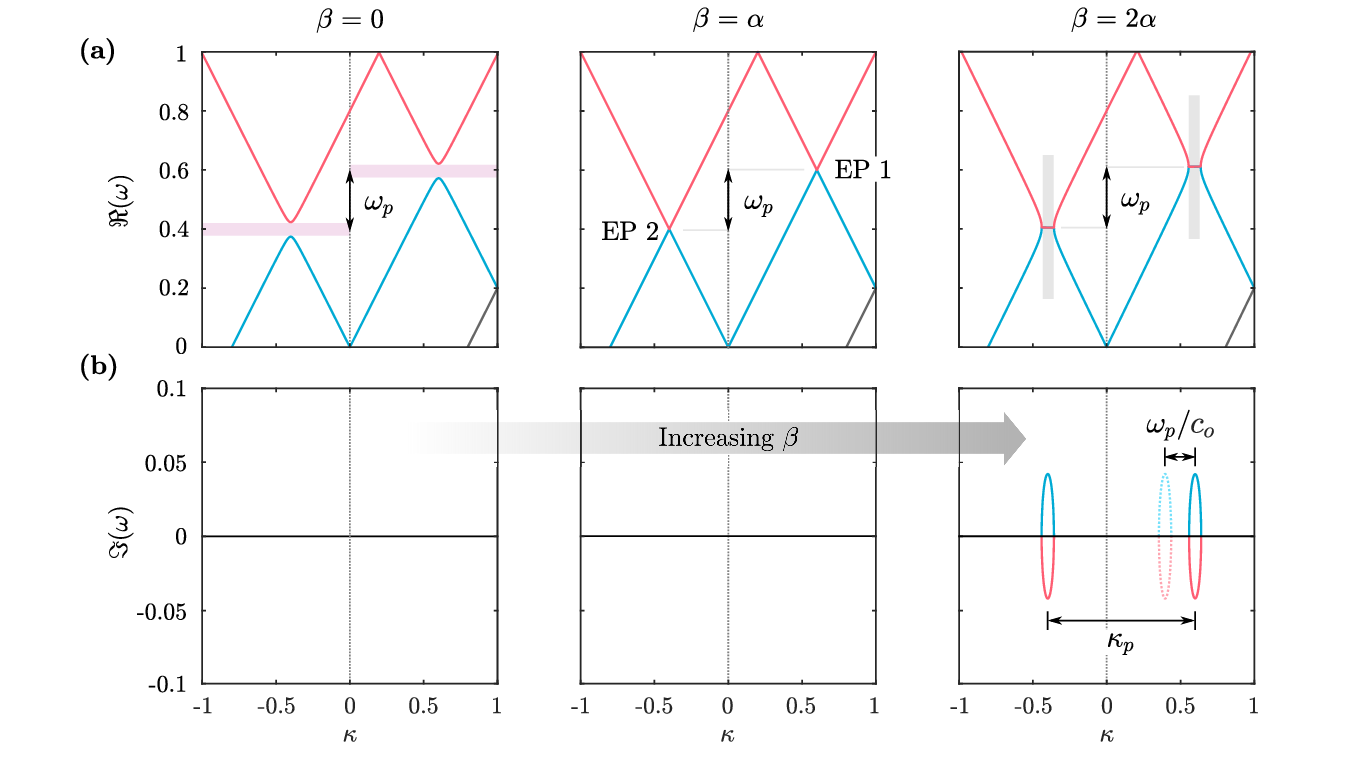}
\caption{Transition from PBGs to $\kappa$ gaps past the $\beta = \alpha$ threshold. (a) The $\Re(\omega)$-$\kappa$ diagrams and (b) $\Im(\omega)$-$\kappa$ diagrams for $\beta = 0$ (left), $\beta = \alpha$ (middle), and $\beta = 2 \alpha$ (right). System parameters used are: $\alpha = 0.2$, $\kappa_p = 1$, $\omega_p = 0.2$, and $E_o = \rho_o = 1$. Vertical and horizontal shifts culminating from band tilting are marked as functions of temporal modulation frequency $\omega_p$, spatial modulation frequency $\kappa_p$, and sonic speed $c_o$.}
\label{fig:transition}
\end{figure}

Figures~\ref{fig:transition}a and b show the $\Re(\omega)$-$\kappa$ and $\Im(\omega)$-$\kappa$ plots of three systems, respectively, corresponding to $\beta=0$, $\beta<\alpha$, and $\beta>\alpha$. In the three cases, $\alpha=0.2$ is chosen to keep the interactions weak and a modulation frequency of $\omega_p=0.2$ is used to break the system's reciprocity. The transition from PBGs ($\beta = 0$) to $\kappa$ gaps ($\beta = 2\alpha$) is obvious and a band structure tilt equal to the value of $\omega_p$ can be observed across all three plots of Fig.~\ref{fig:transition}a. In the leftmost plot, the tilt represents the frequency shift between the PBGs associated with forward and backward waves. In the middle plot, it represents the frequency shift between the two exceptional points, labeled EP 1 and EP 2. In the rightmost plot, it represents the frequency shift between the $\kappa$ gaps associated with forward and backward waves. In addition to the vertical frequency shift, it is important to note that the emergent $\kappa$ gaps in the $\beta=2\alpha$ scenario are also asymmetric with respect to the $\kappa=0$ line and, consequently, exhibit a horizontal shift that can be interestingly quantified in two ways. The small shift between the right $\kappa$ gap and the reflected version of the left $\kappa$ gap (shown via faded colors) is equal to $\omega_p / c_o$, while the overall horizontal distance between the two $\kappa$ gaps is equal to the spatial modulation $\kappa_p$.

\begin{figure}[tbh!]
\centering
\includegraphics[width=0.9\linewidth]{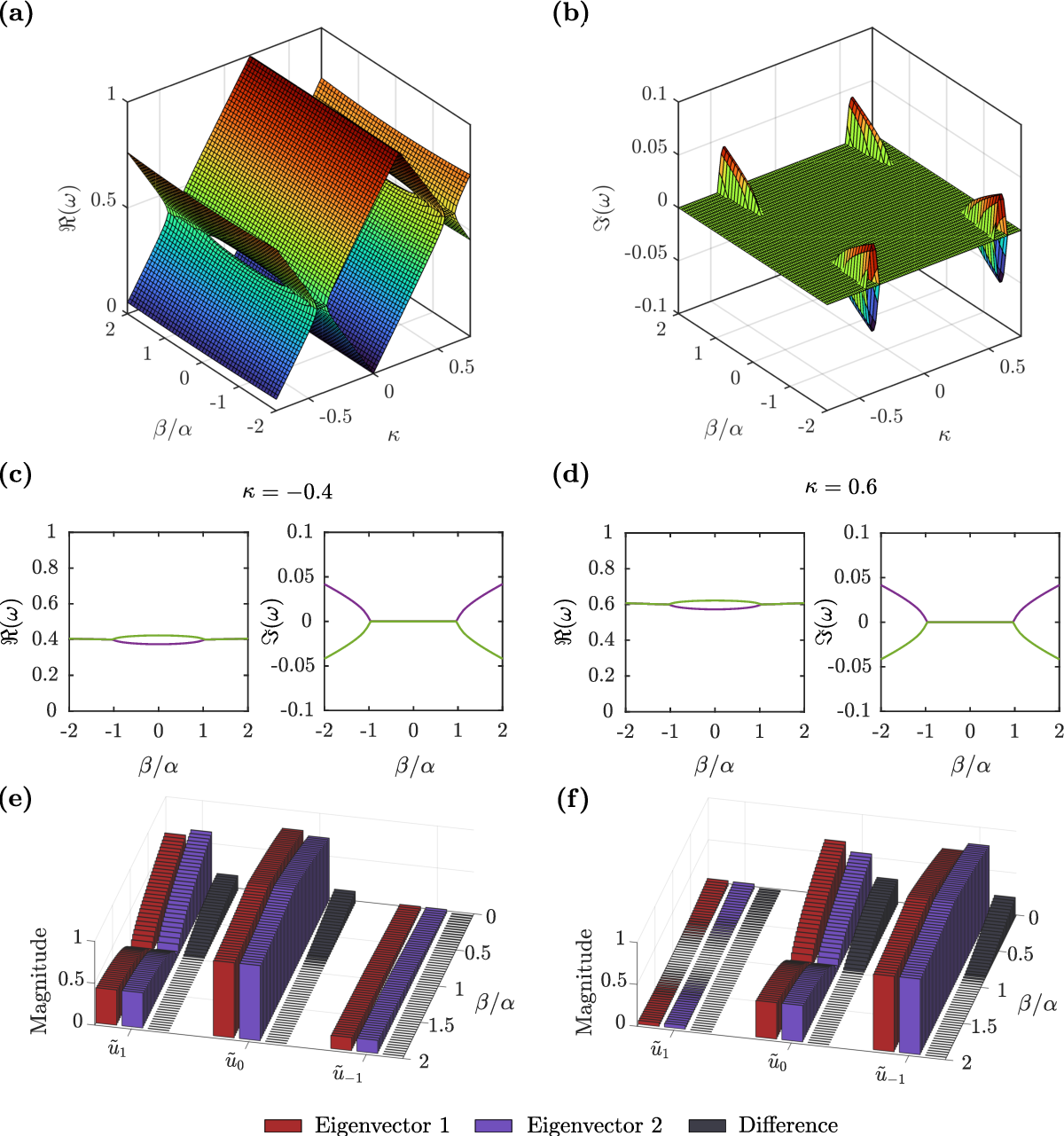}
\caption{Effect of $\beta/\alpha$ on the dispersion behavior, $\kappa$ gaps, and exceptional points. (a) The $\Re(\omega)$-$\kappa$ and (b) $\Im(\omega)$-$\kappa$ relations for different $\beta / \alpha$ values. Exceptional points can be observed at $|\beta / \alpha|= 1$. (c)-(f) Coalescence of eigenvalues and eigenvectors for (c),(e) $\kappa = -0.4$ and (d),(f) $\kappa = 0.6$. Components of the first and second eigenvectors are different for any $\beta/\alpha < 1$ and identical for $\beta/\alpha \geq 1$, as confirmed by the difference bar in both (e) and (f). System parameters used are $\alpha = 0.2$, $\kappa_p = 1$, $\omega_p = 0.2$, and $E_o = \rho_o = 1$.}
\label{fig:coalesce}
\end{figure}

Figure~\ref{fig:coalesce} shows the evolution of the dispersion behavior of the system as the ratio $\beta / \alpha$ varies between $-2$ and $2$. Of interest is the case where $\beta=\alpha$, and $\mathbf{C}_{\lfloor d=1 \rfloor}$ simplifies to
\begin{align}
\label{eq:exp_point}
\mathbf{C}_{\lfloor d=1 \rfloor} \bigg\rvert_{\beta = \alpha} =
   \begin{bmatrix}
    \omega_p^2 - c_o^2 (\kappa-\kappa_p)^2 & -c_o^2 \kappa (\kappa-\kappa_p) \alpha & 0\\
    0 & -c_o^2 \kappa^2 & -c_o^2 \kappa (\kappa+\kappa_p) \alpha\\
    0 & 0 &  \omega_p^2 - c_o^2 (\kappa+\kappa_p)^2 \\
    \end{bmatrix}.
\end{align}
For a certain set of $\kappa$ values, the matrix given by Eq.~(\ref{eq:exp_point}) becomes a defective stiffness matrix that cannot be diagonalized. In such a case, the corresponding system would not support a complete basis of eigenvectors and its algebraic multiplicity exceeds its geometric multiplicity. This phenomena is attributed to exceptional points in the parameter space that, in this problem, appear at the $|\beta/\alpha| = 1$ point for $\kappa$ values at the boundaries of the Brillouin zone. As shown in Figs.~\ref{fig:coalesce}c and d, the eigenvalues of the system coalesce at this point and become complex conjugates beyond it. Figures~\ref{fig:coalesce}e and f show that the two eigenvectors, represented by $\mathbf{\tilde{u}}_{\lfloor d=1 \rfloor}$, become perfectly identical at the same point, which is verified by their difference being zero for any $|\beta/\alpha| \geq 1$, further confirming that the dynamics of the system significantly change before and after each exceptional point.

\vspace{0.25cm}

\newpage

\begin{figure}[h!]
\centering
\includegraphics[width=1\linewidth]{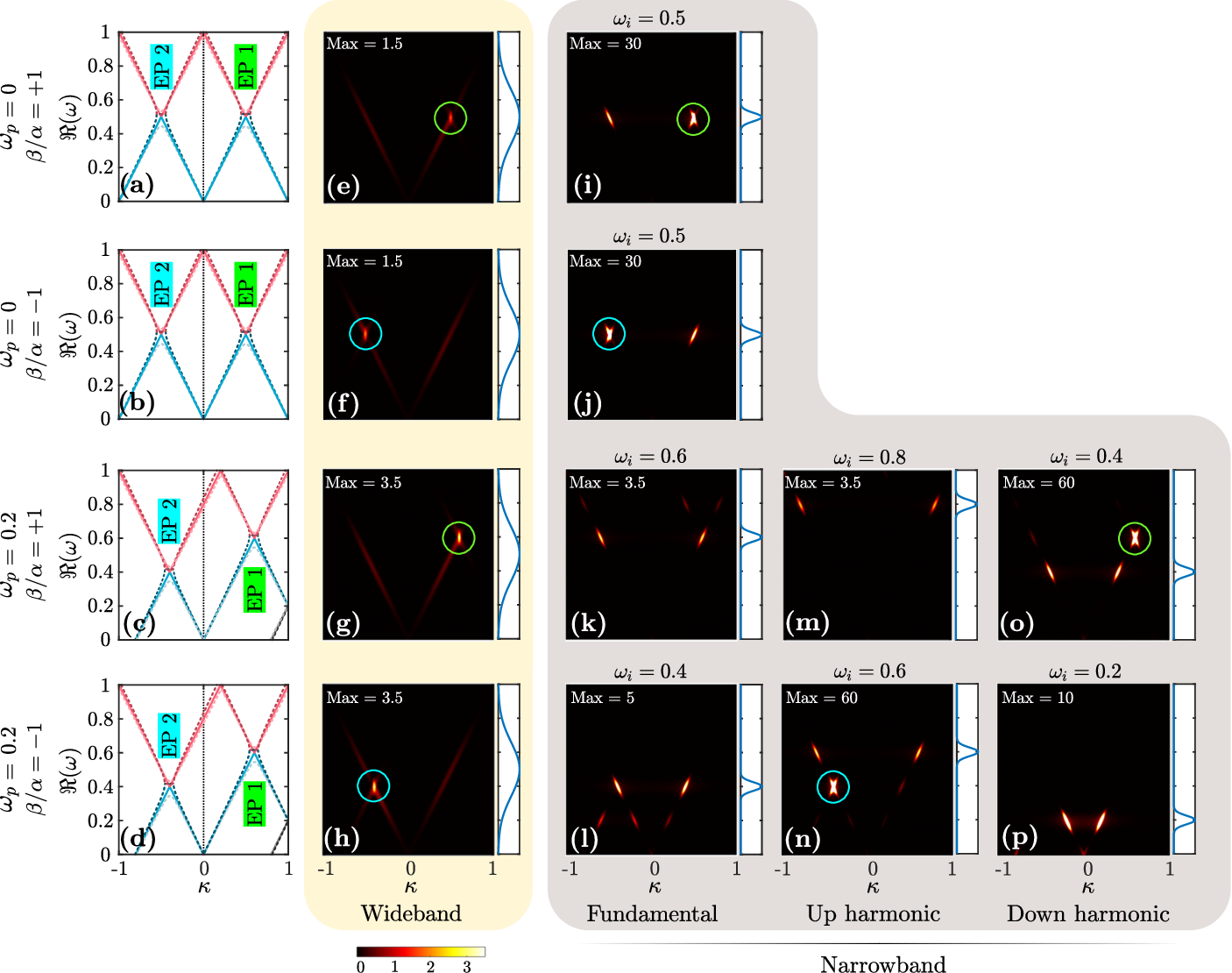}
\caption{Directional amplification at EPs. (a)-(d) The $\Re(\omega)$-$\kappa$ dispersion diagrams highlighting the locations of EP 1 and EP 2 for (a) $\omega_p = 0$ and $\beta/\alpha=+1$, (b) $\omega_p = 0$ and $\beta/\alpha=-1$, (c) $\omega_p = 0.2$ and $\beta/\alpha=+1$, and (d) $\omega_p = 0.2$ and $\beta/\alpha=-1$. Solid lines in each figure show the dispersion bands at $|\beta / \alpha|= 1$ while dashed lines show the transition from $|\beta / \alpha| < 1$ to $|\beta / \alpha| > 1$. (e)-(h) Numerically constructed dispersion contours from the system's response to a wideband excitation. Excitation waveform is provided alongside the figure for visualization. Amplification at the location of EP 1 or EP 2 is marked using green or blue circles, respectively. (i)-(l) Numerically constructed dispersion contours from the system's response to a narrowband excitation targeting the frequency of the amplified EP shown in (e)-(h), i.e., $\omega_i = \omega_{\text{EP/amp}}$. (m)-(n) Numerically constructed dispersion contours from the system's response to a narrowband excitation targeting the first up harmonic of the amplified EP frequency shown in (g)-(h), i.e., $\omega_i = \omega_{\text{EP/amp}}+\omega_p$. (o)-(p) Numerically constructed dispersion contours from the system's response to a narrowband excitation targeting the first down harmonic of the amplified EP frequency shown in (g)-(h), i.e., $\omega_i = \omega_{\text{EP/amp}}-\omega_p$. For comparison purposes, the color bar is kept unchanged throughout the figure while the maximum amplitude is listed in the top left corner of each case. System parameters used are: $\alpha = 0.2$, $\kappa_p = 1$, and $E_o = \rho_o = 1$.}
\label{fig:EP}
\end{figure}

Next, we investigate the amplification behavior associated with exceptional points at $|\beta / \alpha| = 1$ in the presence and absence of a temporal modulation. The solid lines in Fig.~\ref{fig:EP}a and b show the system's dispersion bands when $\beta / \alpha=\pm 1$ in the absence of a temporal modulation, i.e., $\omega_p = 0$. The bands are color coded in the same way they have been throughout this paper, and the two exceptional points, EP 1 and EP 2, are indicated on the figure. In this case, EP 1 and EP 2 are symmetric with respect to the $\kappa=0$ axis and share the same frequency, $\omega_{\text{EP}} = 0.5$. The dashed lines in the same figure show the transition of the dispersion diagram from $\beta / \alpha < 1$ to $\beta / \alpha > 1$, namely the changeover from PBGs to $\kappa$ gaps. Figure~\ref{fig:EP}c shows the same set of dispersion lines but when a temporal modulation of $\omega_p=0.2$ is introduced. Finally, Fig.~\ref{fig:EP}d shows the behavior of the system when $\beta / \alpha=-1$ and $\omega_p=0.2$. Owing to the value of $\omega_p$, EP 1 and EP 2 in both Figs.~\ref{fig:EP}c and d take place at $\omega_{\text{EP}} = 0.6$ and $\omega_{\text{EP}} = 0.4$, respectively. Figures~\ref{fig:EP}e through p show the numerically reconstructed dispersion surfaces that are obtained using the ``General Form PDE" module of COMSOL Multiphysics, with Eq.~(\ref{eq:vib}) being directly used to formulate the model. The geometry is constructed using 2000 discrete points along a 1D array with a resolution of 10 points per wavelength, and a zero flux boundary condition is applied to both ends. A Gaussian excitation of the form $f(t) = e^{ (t_0-t)^2/2 \sigma} \sin{(\omega_{i} t)}$ is applied using the flux (source) boundary condition at the middle of the model and 2D FFTs are performed on the time-domain fields to obtain the dispersion data. For the wideband excitations, the following excitation parameters are chosen: $\sigma = 1$, $t_0 = 20$, and $\omega_{i} = 0.5$. To impose narrowband excitations, $\sigma$ is set to $40$ and $\omega_{i}$ is tuned to the center frequency of interest, as indicated in the individual figures. For convenience, the shape of the exciting waveform is plotted alongside each simulation. To avoid reflections, the simulations are only performed up to the time required for the waves to reach the boundaries.

By inspecting the wideband excitation cases in Figs.~\ref{fig:EP}e-h, significant amplification can be observed in each of the four cases at one of the two aforementioned EPs. The sign of the $\beta / \alpha$ ratio dictates which EP becomes amplified. The amplification of EP 1 corresponds to $\beta / \alpha = +1$ (e.g., Figs.~\ref{fig:EP}e and g) while the amplification of EP 2 corresponds to $\beta / \alpha = -1$ (e.g., Fig.~\ref{fig:EP}f and h). We henceforth refer to the frequency of the amplified EP as $\omega_{\text{EP/amp}}$. To further investigate the behavior of these EPs, a narrowband excitation targeting the frequency of the amplified EP, i.e., $\omega_i = \omega_{\text{EP/amp}}$, is applied to each of the four cases. In the absence of $\omega_p$, the narrowband excitations in Fig.~\ref{fig:EP}i and j generate the same amplification as that of the wideband excitations shown in Fig.~\ref{fig:EP}e and f, respectively. However, in the presence of a temporal modulation of $\omega_p = 0.2$, neither one of the narrowband excitations used in Figs.~\ref{fig:EP}k and l is able to generate the same amplification achieved by their respective wideband excitations in Figs.~\ref{fig:EP}g and h. Given the influence of temporal modulation on the underlying system dynamics, exciting the structure at different up and down harmonics of the amplified EP frequency becomes relevant \cite{moghaddaszadeh2021nonreciprocal}. To this end, a narrowband excitation corresponding to $\omega_i = \omega_{\text{EP/amp}}+\omega_p$ is imposed in Figs.~\ref{fig:EP}m and n, and a narrowband excitation corresponding to $\omega_i = \omega_{\text{EP/amp}}-\omega_p$ is imposed in Figs.~\ref{fig:EP}o and p. The results show that no amplification takes place in either Fig.~\ref{fig:EP}m and p, but does take place in both Fig.~\ref{fig:EP}o and n at the same EP that is shown to be amplifiable using a wideband excitation. As such, the following conclusions can be made.

\begin{enumerate}
    \item In the absence of a temporal modulation (i.e., $\omega_p = 0$), both wide and narrowband excitations with $\omega_i = \omega_{\text{EP}}$ will trigger an amplification at the EPs whose location has the same sign for $\kappa$ and $\beta / \alpha$. Here, EP 1 is amplified when $\beta / \alpha = +1$, while EP 2 is amplified when $\beta / \alpha = -1$.
    \item In the presence of a temporal modulation (i.e., $\omega_p \neq 0)$ the following statements hold:
    
    \begin{enumerate}
        \item A wideband excitation will result in amplification at a single EP location, $\omega_{\text{EP/amp}}$. However, a narrowband excitation of $\omega_i = \omega_{\text{EP/amp}}$ will fail to reproduce such amplification. 
        \item For $\beta / \alpha = \pm 1$ and an amplifiable EP at $\omega_{\text{EP/amp}}$, a narrowband excitation of $\omega_i = \omega_{\text{EP/amp}} \mp \omega_p$ is needed to activate the EP amplification. 
    \end{enumerate}
\end{enumerate}

\begin{figure}[h!]
\centering
\includegraphics[width=\linewidth]{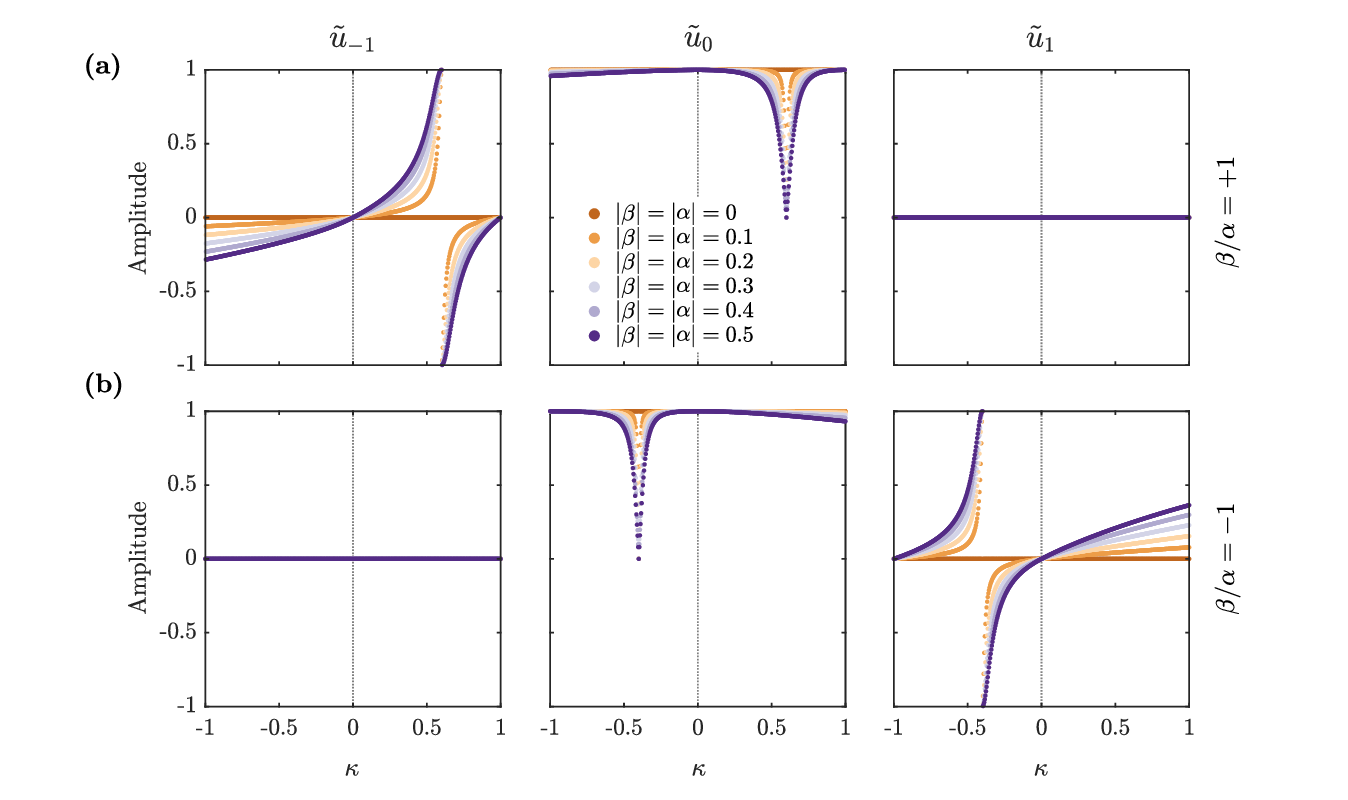}
\caption{Variation of the different components of the first eigenvector as a function of $\kappa$ for different magnitudes of $|\beta| = |\alpha|$ with (a) $\beta / \alpha = +1$ and (b) $\beta / \alpha = -1$. Here $\tilde{u}_{-1}$ and $\tilde{u}_{1}$ denote the down and up harmonic components of the eigenvector, respectively, while $\tilde{u}_{0}$ represents the fundamental component. System parameters used are $\alpha = 0.2$, $\kappa_p = 1$, $\omega_p = 0.2$, and $E_o = \rho_o = 1$.}
\label{fig:eigenvectors}
\end{figure}

To further elaborate, consider the analysis presented in Fig.~\ref{fig:eigenvectors} where different components of the first eigenvector are plotted for several scenarios, all sharing the following feature: $|\beta| = |\alpha|$. In Fig.~\ref{fig:eigenvectors}a, where $\beta / \alpha = +1$, a noticeable drop in the amplitude of $\tilde{u}_{0}$ is observed at $\kappa = 0.6$. Leading up to the same value of $\kappa$, a rise in the amplitude of $\tilde{u}_{-1}$ can also be observed. This is indicative of an amplification taking place at EP 1 (here at $\kappa = 0.6$) corresponding to $\beta / \alpha = +1$, which is consistent with the results obtained from Fig.~\ref{fig:EP}. Furthermore, since $\tilde{u}_{-1}$ represents the down harmonic component of the eigenvector as implied by the PWE solution in Eq.~(\ref{eq:solutionl}), this also explains why the EP amplification via a narrowband excitation is only attainable by targeting the down harmonic of $\omega_{\text{EP/amp}}$ in Fig.~\ref{fig:EP}o. Similarly, in Fig.~\ref{fig:eigenvectors}b, where $\beta / \alpha = -1$, a noticeable drop in the amplitude of $\tilde{u}_{0}$ is observed at $\kappa = -0.4$. Leading up to the same value of $\kappa$, a rise in the amplitude of $\tilde{u}_{+1}$ can also be observed. This is indicative of an amplification taking place at EP 2 (here at $\kappa = -0.4$) corresponding to $\beta / \alpha = -1$, which is consistent with the results obtained from Fig.~\ref{fig:EP}. Furthermore, since $\tilde{u}_{+1}$ represents the up harmonic component of the eigenvector, this explains why the EP amplification via a narrowband excitation is only attainable by targeting the up harmonic of $\omega_{\text{EP/amp}}$ in Fig.~\ref{fig:EP}n. Finally, it is worth noting that the location of the EP remains unaltered regardless of the equal magnitudes of $\alpha$ and $\beta$, as can be inferred from the nonchanging locations of the amplitude changes in $\tilde{u}_{0}$, $\tilde{u}_{-1}$, and $\tilde{u}_{1}$ in Fig.~\ref{fig:eigenvectors}. However, the distribution of energy between different harmonics changes depending on the magnitude of $\alpha$ or $\beta$. Needless to mention, the observations made in this section can be generalized for all higher-order harmonics (i.e., $\omega_\text{EP/amp} \pm n \omega_p$ for $n \in \mathbb{Z}$ and $n \neq 0$).

\section{Conclusions}
\label{Sec:con}
In this work we investigate the wave dispersion mechanics of a one-dimensional elastic medium with a stiffness profile that is modulated using real (even) or imaginary (odd) space-time waveforms corresponding to different modulation speeds, each revealing intriguing features across the frequency and wave-number axes, which can be summarized as follows. For even, real spatiotemporal modulations, either nonreciprocal PBGs or asymmetric $\kappa$ gaps form depending on whether the modulation speed ratio, $\nu$, is smaller or greater than one. At the limit states of $\omega_p = 0$ or $\kappa_p = 0$, reciprocal PBGs or symmetric $\kappa$ gaps will be observed, respectively. For odd, imaginary spatiotemporal modulations, asymmetric odd-ordered $\kappa$ gaps form when $\nu < 1$, while a combination of nonreciprocal PBGs and asymmetric even-ordered $\kappa$ gaps form when $\nu > 1$. At the limit states of $\omega_p = 0$ or $\kappa_p = 0$, symmetric odd-ordered $\kappa$ gaps or a combination of reciprocal PBGs and symmetric even-ordered $\kappa$ gaps will be observed, respectively. The above takeaways are qualitatively captured by eight distinct examples, labeled cases 1 through 8, which cover all the possible spatial, temporal, and spatiotemporal modulation profiles in both real and imaginary forms. The resultant behaviors of the systems undergoing these modulations are summarized in Table~\ref{table:1}.

Beyond the eight individually examined cases, the analysis of complex modulations with simultaneous real and imaginary components show a large dependence on the absolute amplitude ratio $|\beta / \alpha|$, which reveals the following.

\newpage
 
\begin{enumerate}
    \item For a generalized complex stiffness modulation of the form $E(x,t) = E_o [1 + \alpha \cos(\omega_p t-\kappa_p x) + \ii \beta \sin(\omega_p t-\kappa_p x)]$, the dispersion behavior approaches that of a system with purely real modulations when $|\beta / \alpha| < 1$, and that of a system with purely imaginary modulations when $|\beta / \alpha| > 1$.
    
    \item The $|\beta / \alpha| =  1$ system gives rise to a series of EPs for both right- and left-going waves. Directional amplification will take place at the former EPs for positive values of $\beta / \alpha$, and for the latter EPs for negative values of $\beta / \alpha$. The frequency of the amplifiable EP is denoted $\omega_{\text{EP/amp}}$.
\end{enumerate}
    
Finally, unlike time-invariant non-Hermitian systems where EP amplification can simply be triggered with a narrow excitation of $\omega_{\text{EP/amp}}$, the amplifiable EP in a non-Hermitian system with a temporal modulation $\omega_p \neq 0$ can be realized by exciting several harmonics of the amplifiable EP frequency, specifically $\omega_{\text{EP/amp}} \mp n \omega_p$ for $\beta / \alpha = \pm 1$ (with $n \in \mathbb{Z}$ and $n \neq 0$); a feature that stems directly from the properties of the different components of the system eigenvectors. From a practical standpoint, the ability to trigger amplification at distinct frequencies, coupled with the significantly enhanced sensitivity to perturbations in the vicinity of EPs, can be extremely valuable given the rapidly increasing use of EPs in optical and acoustic sensing devices \cite{wiersig2020review,rosa2021exceptional}. In the presence of temporal modulations, EP-based sensors can therefore operate within a wider bandwidth since the amplification can be tuned to target multiple frequencies of interest spanning a broader range. 

\begin{table}[h!]
\caption{Summary of key features accompanying different stiffness modulation forms of a one-dimensional elastic medium. The abbreviations R, NR, S, and AS denote reciprocal, nonreciprocal, symmetric, and asymmetric, respectively; ``/s" denotes a shifting PBG at higher values of $\kappa$, as demonstrated in Fig.~\ref{fig:PBG_shrink}.}
\vspace{0.2cm}
\centering
\begin{tabular*}{\textwidth}{c @{\extracolsep{\fill}} lcccc}
\hline \\ [-0.9em]
\multicolumn{1}{c}{\textbf{Case}} &
\multicolumn{1}{l}{\textbf{Modulation}} &
\multicolumn{1}{c}{\textbf{PBGs}}&
\multicolumn{3}{c}{\textbf{$\kappa$ gaps}}
\\
\cline{4-6}
\rule{0pt}{9pt}& & & Symmetry & Order & Level with increasing gap order \\
[0.2em] 
\hline \\ [-0.8em]
1 & $\cos(-\kappa_p x)$ & R & -& - & - \\ 
2 & $\cos(\omega_p t)$ & - & S & All & Decreasing\\
3 & $\cos(\omega_p t-\kappa_p x)$; $\nu<1$ & NR & - & - & - \\
4 & $\cos(\omega_p t-\kappa_p x)$; $\nu>1$ & - & AS & All & Decreasing\\
5 & $\ii \sin(-\kappa_p x)$ & - & S & Odd & Constant \\
6 & $\ii \sin(\omega_p t)$ & R/s & S & Even & Decreasing  \\
7 & $\ii \sin(\omega_p t-\kappa_p x)$; $\nu<1$ & - & AS & Odd & Constant \\
8 & $\ii \sin(\omega_p t-\kappa_p x)$; $\nu>1$ & NR/s & AS & Even & Decreasing  \\ [0.2em]
\hline
\end{tabular*}
\label{table:1}
\end{table}

\section*{Acknowledgments}
The authors acknowledge support from the US National Science Foundation through Grant No. 1847254 (CAREER), as well as support from the University at Buffalo (SUNY) through the Buffalo Blue Sky Program.


\bibliography{references}

\begin{thebibliography}{56}%
\makeatletter
\providecommand \@ifxundefined [1]{%
 \@ifx{#1\undefined}
}%
\providecommand \@ifnum [1]{%
 \ifnum #1\expandafter \@firstoftwo
 \else \expandafter \@secondoftwo
 \fi
}%
\providecommand \@ifx [1]{%
 \ifx #1\expandafter \@firstoftwo
 \else \expandafter \@secondoftwo
 \fi
}%
\providecommand \natexlab [1]{#1}%
\providecommand \enquote  [1]{``#1''}%
\providecommand \bibnamefont  [1]{#1}%
\providecommand \bibfnamefont [1]{#1}%
\providecommand \citenamefont [1]{#1}%
\providecommand \href@noop [0]{\@secondoftwo}%
\providecommand \href [0]{\begingroup \@sanitize@url \@href}%
\providecommand \@href[1]{\@@startlink{#1}\@@href}%
\providecommand \@@href[1]{\endgroup#1\@@endlink}%
\providecommand \@sanitize@url [0]{\catcode `\\12\catcode `\$12\catcode
  `\&12\catcode `\#12\catcode `\^12\catcode `\_12\catcode `\%12\relax}%
\providecommand \@@startlink[1]{}%
\providecommand \@@endlink[0]{}%
\providecommand \url  [0]{\begingroup\@sanitize@url \@url }%
\providecommand \@url [1]{\endgroup\@href {#1}{\urlprefix }}%
\providecommand \urlprefix  [0]{URL }%
\providecommand \Eprint [0]{\href }%
\providecommand \doibase [0]{https://doi.org/}%
\providecommand \selectlanguage [0]{\@gobble}%
\providecommand \bibinfo  [0]{\@secondoftwo}%
\providecommand \bibfield  [0]{\@secondoftwo}%
\providecommand \translation [1]{[#1]}%
\providecommand \BibitemOpen [0]{}%
\providecommand \bibitemStop [0]{}%
\providecommand \bibitemNoStop [0]{.\EOS\space}%
\providecommand \EOS [0]{\spacefactor3000\relax}%
\providecommand \BibitemShut  [1]{\csname bibitem#1\endcsname}%
\let\auto@bib@innerbib\@empty
\bibitem [{\citenamefont {Nassar}\ \emph {et~al.}(2020)\citenamefont {Nassar},
  \citenamefont {Yousefzadeh}, \citenamefont {Fleury}, \citenamefont {Ruzzene},
  \citenamefont {Al{\`u}}, \citenamefont {Daraio}, \citenamefont {Norris},
  \citenamefont {Huang},\ and\ \citenamefont
  {Haberman}}]{nassar2020nonreciprocity}%
  \BibitemOpen
  \bibfield  {author} {\bibinfo {author} {\bibfnamefont {H.}~\bibnamefont
  {Nassar}}, \bibinfo {author} {\bibfnamefont {B.}~\bibnamefont {Yousefzadeh}},
  \bibinfo {author} {\bibfnamefont {R.}~\bibnamefont {Fleury}}, \bibinfo
  {author} {\bibfnamefont {M.}~\bibnamefont {Ruzzene}}, \bibinfo {author}
  {\bibfnamefont {A.}~\bibnamefont {Al{\`u}}}, \bibinfo {author} {\bibfnamefont
  {C.}~\bibnamefont {Daraio}}, \bibinfo {author} {\bibfnamefont {A.~N.}\
  \bibnamefont {Norris}}, \bibinfo {author} {\bibfnamefont {G.}~\bibnamefont
  {Huang}},\ and\ \bibinfo {author} {\bibfnamefont {M.~R.}\ \bibnamefont
  {Haberman}},\ }\bibfield  {title} {\bibinfo {title} {Nonreciprocity in
  acoustic and elastic materials},\ }\href@noop {} {\bibfield  {journal}
  {\bibinfo  {journal} {Nat. Rev. Mater.}\ }\textbf {\bibinfo {volume} {5}},\
  \bibinfo {pages} {667} (\bibinfo {year} {2020})}\BibitemShut {NoStop}%
\bibitem [{\citenamefont {Al~Ba'ba'a}\ \emph {et~al.}(2019)\citenamefont
  {Al~Ba'ba'a}, \citenamefont {Nouh},\ and\ \citenamefont
  {Singh}}]{al2019dispersion}%
  \BibitemOpen
  \bibfield  {author} {\bibinfo {author} {\bibfnamefont {H.}~\bibnamefont
  {Al~Ba'ba'a}}, \bibinfo {author} {\bibfnamefont {M.}~\bibnamefont {Nouh}},\
  and\ \bibinfo {author} {\bibfnamefont {T.}~\bibnamefont {Singh}},\ }\bibfield
   {title} {\bibinfo {title} {Dispersion and topological characteristics of
  permutative polyatomic phononic crystals},\ }\href@noop {} {\bibfield
  {journal} {\bibinfo  {journal} {Proc. Royal Soc. A}\ }\textbf {\bibinfo
  {volume} {475}},\ \bibinfo {pages} {20190022} (\bibinfo {year}
  {2019})}\BibitemShut {NoStop}%
\bibitem [{\citenamefont {Su}\ \emph {et~al.}(2012)\citenamefont {Su},
  \citenamefont {Gao},\ and\ \citenamefont {Zhou}}]{su2012influence}%
  \BibitemOpen
  \bibfield  {author} {\bibinfo {author} {\bibfnamefont {X.}~\bibnamefont
  {Su}}, \bibinfo {author} {\bibfnamefont {Y.}~\bibnamefont {Gao}},\ and\
  \bibinfo {author} {\bibfnamefont {Y.}~\bibnamefont {Zhou}},\ }\bibfield
  {title} {\bibinfo {title} {The influence of material properties on the
  elastic band structures of one-dimensional functionally graded phononic
  crystals},\ }\href@noop {} {\bibfield  {journal} {\bibinfo  {journal} {J.
  Appl. Phys.}\ }\textbf {\bibinfo {volume} {112}},\ \bibinfo {pages} {123503}
  (\bibinfo {year} {2012})}\BibitemShut {NoStop}%
\bibitem [{\citenamefont {Ansari}\ and\ \citenamefont
  {Karami}(2017)}]{ansari2017analyzing}%
  \BibitemOpen
  \bibfield  {author} {\bibinfo {author} {\bibfnamefont {M.~H.}\ \bibnamefont
  {Ansari}}\ and\ \bibinfo {author} {\bibfnamefont {M.~A.}\ \bibnamefont
  {Karami}},\ }\bibfield  {title} {\bibinfo {title} {Analyzing the frequency
  band gap in functionally graded materials with harmonically varying material
  properties},\ }\href@noop {} {\bibfield  {journal} {\bibinfo  {journal}
  {Proc. SPIE}\ ,\ \bibinfo {pages} {101701}} (\bibinfo {year}
  {2017})}\BibitemShut {NoStop}%
\bibitem [{\citenamefont {Trainiti}\ \emph {et~al.}(2019)\citenamefont
  {Trainiti}, \citenamefont {Xia}, \citenamefont {Marconi}, \citenamefont
  {Cazzulani}, \citenamefont {Erturk},\ and\ \citenamefont
  {Ruzzene}}]{trainiti2019time}%
  \BibitemOpen
  \bibfield  {author} {\bibinfo {author} {\bibfnamefont {G.}~\bibnamefont
  {Trainiti}}, \bibinfo {author} {\bibfnamefont {Y.}~\bibnamefont {Xia}},
  \bibinfo {author} {\bibfnamefont {J.}~\bibnamefont {Marconi}}, \bibinfo
  {author} {\bibfnamefont {G.}~\bibnamefont {Cazzulani}}, \bibinfo {author}
  {\bibfnamefont {A.}~\bibnamefont {Erturk}},\ and\ \bibinfo {author}
  {\bibfnamefont {M.}~\bibnamefont {Ruzzene}},\ }\bibfield  {title} {\bibinfo
  {title} {Time-periodic stiffness modulation in elastic metamaterials for
  selective wave filtering: Theory and experiment},\ }\href@noop {} {\bibfield
  {journal} {\bibinfo  {journal} {Phys. Rev. Lett.}\ }\textbf {\bibinfo
  {volume} {122}},\ \bibinfo {pages} {124301} (\bibinfo {year}
  {2019})}\BibitemShut {NoStop}%
\bibitem [{\citenamefont {Vila}\ \emph {et~al.}(2017)\citenamefont {Vila},
  \citenamefont {Pal}, \citenamefont {Ruzzene},\ and\ \citenamefont
  {Trainiti}}]{Vila2017363}%
  \BibitemOpen
  \bibfield  {author} {\bibinfo {author} {\bibfnamefont {J.}~\bibnamefont
  {Vila}}, \bibinfo {author} {\bibfnamefont {R.~K.}\ \bibnamefont {Pal}},
  \bibinfo {author} {\bibfnamefont {M.}~\bibnamefont {Ruzzene}},\ and\ \bibinfo
  {author} {\bibfnamefont {G.}~\bibnamefont {Trainiti}},\ }\bibfield  {title}
  {\bibinfo {title} {A bloch-based procedure for dispersion analysis of
  lattices with periodic time-varying properties},\ }\href@noop {} {\bibfield
  {journal} {\bibinfo  {journal} {J. Sound Vib.}\ }\textbf {\bibinfo {volume}
  {406}},\ \bibinfo {pages} {363 } (\bibinfo {year} {2017})}\BibitemShut
  {NoStop}%
\bibitem [{\citenamefont {Nassar}\ \emph
  {et~al.}(2017{\natexlab{a}})\citenamefont {Nassar}, \citenamefont {Xu},
  \citenamefont {Norris},\ and\ \citenamefont {Huang}}]{Nassar201710}%
  \BibitemOpen
  \bibfield  {author} {\bibinfo {author} {\bibfnamefont {H.}~\bibnamefont
  {Nassar}}, \bibinfo {author} {\bibfnamefont {X.}~\bibnamefont {Xu}}, \bibinfo
  {author} {\bibfnamefont {A.}~\bibnamefont {Norris}},\ and\ \bibinfo {author}
  {\bibfnamefont {G.}~\bibnamefont {Huang}},\ }\bibfield  {title} {\bibinfo
  {title} {Modulated phononic crystals: Non-reciprocal wave propagation and
  willis materials},\ }\href@noop {} {\bibfield  {journal} {\bibinfo  {journal}
  {J. Mech. Phys. Solids}\ }\textbf {\bibinfo {volume} {101}},\ \bibinfo
  {pages} {10 } (\bibinfo {year} {2017}{\natexlab{a}})}\BibitemShut {NoStop}%
\bibitem [{\citenamefont {Attarzadeh}\ and\ \citenamefont
  {Nouh}(2018)}]{attarzadeh2018non}%
  \BibitemOpen
  \bibfield  {author} {\bibinfo {author} {\bibfnamefont {M.~A.}\ \bibnamefont
  {Attarzadeh}}\ and\ \bibinfo {author} {\bibfnamefont {M.}~\bibnamefont
  {Nouh}},\ }\bibfield  {title} {\bibinfo {title} {Non-reciprocal elastic wave
  propagation in 2d phononic membranes with spatiotemporally varying material
  properties},\ }\href@noop {} {\bibfield  {journal} {\bibinfo  {journal} {J.
  Sound Vib.}\ }\textbf {\bibinfo {volume} {422}},\ \bibinfo {pages} {264}
  (\bibinfo {year} {2018})}\BibitemShut {NoStop}%
\bibitem [{\citenamefont {Trainiti}\ and\ \citenamefont
  {Ruzzene}(2016)}]{trainiti2016non}%
  \BibitemOpen
  \bibfield  {author} {\bibinfo {author} {\bibfnamefont {G.}~\bibnamefont
  {Trainiti}}\ and\ \bibinfo {author} {\bibfnamefont {M.}~\bibnamefont
  {Ruzzene}},\ }\bibfield  {title} {\bibinfo {title} {Non-reciprocal elastic
  wave propagation in spatiotemporal periodic structures},\ }\href@noop {}
  {\bibfield  {journal} {\bibinfo  {journal} {New J. Phys.}\ }\textbf {\bibinfo
  {volume} {18}},\ \bibinfo {pages} {083047} (\bibinfo {year}
  {2016})}\BibitemShut {NoStop}%
\bibitem [{\citenamefont {Attarzadeh}\ \emph {et~al.}(2018)\citenamefont
  {Attarzadeh}, \citenamefont {Al~Ba’ba’a},\ and\ \citenamefont
  {Nouh}}]{attarzadeh2018wave}%
  \BibitemOpen
  \bibfield  {author} {\bibinfo {author} {\bibfnamefont {M.~A.}\ \bibnamefont
  {Attarzadeh}}, \bibinfo {author} {\bibfnamefont {H.}~\bibnamefont
  {Al~Ba’ba’a}},\ and\ \bibinfo {author} {\bibfnamefont {M.}~\bibnamefont
  {Nouh}},\ }\bibfield  {title} {\bibinfo {title} {On the wave dispersion and
  non-reciprocal power flow in space-time traveling acoustic metamaterials},\
  }\href@noop {} {\bibfield  {journal} {\bibinfo  {journal} {Appl. Acoust.}\
  }\textbf {\bibinfo {volume} {133}},\ \bibinfo {pages} {210} (\bibinfo {year}
  {2018})}\BibitemShut {NoStop}%
\bibitem [{\citenamefont {Attarzadeh}\ \emph {et~al.}(2020)\citenamefont
  {Attarzadeh}, \citenamefont {Callanan},\ and\ \citenamefont
  {Nouh}}]{attarzadeh2020experimental}%
  \BibitemOpen
  \bibfield  {author} {\bibinfo {author} {\bibfnamefont {M.~A.}\ \bibnamefont
  {Attarzadeh}}, \bibinfo {author} {\bibfnamefont {J.}~\bibnamefont
  {Callanan}},\ and\ \bibinfo {author} {\bibfnamefont {M.}~\bibnamefont
  {Nouh}},\ }\bibfield  {title} {\bibinfo {title} {Experimental observation of
  nonreciprocal waves in a resonant metamaterial beam},\ }\href@noop {}
  {\bibfield  {journal} {\bibinfo  {journal} {Phys. Rev. Appl.}\ }\textbf
  {\bibinfo {volume} {13}},\ \bibinfo {pages} {021001} (\bibinfo {year}
  {2020})}\BibitemShut {NoStop}%
\bibitem [{\citenamefont {Nassar}\ \emph
  {et~al.}(2017{\natexlab{b}})\citenamefont {Nassar}, \citenamefont {Chen},
  \citenamefont {Norris}, \citenamefont {Haberman},\ and\ \citenamefont
  {Huang}}]{nassar2017non}%
  \BibitemOpen
  \bibfield  {author} {\bibinfo {author} {\bibfnamefont {H.}~\bibnamefont
  {Nassar}}, \bibinfo {author} {\bibfnamefont {H.}~\bibnamefont {Chen}},
  \bibinfo {author} {\bibfnamefont {A.}~\bibnamefont {Norris}}, \bibinfo
  {author} {\bibfnamefont {M.}~\bibnamefont {Haberman}},\ and\ \bibinfo
  {author} {\bibfnamefont {G.}~\bibnamefont {Huang}},\ }\bibfield  {title}
  {\bibinfo {title} {Non-reciprocal wave propagation in modulated elastic
  metamaterials},\ }\href@noop {} {\bibfield  {journal} {\bibinfo  {journal}
  {Proc. Royal Soc. A}\ }\textbf {\bibinfo {volume} {473}},\ \bibinfo {pages}
  {20170188} (\bibinfo {year} {2017}{\natexlab{b}})}\BibitemShut {NoStop}%
\bibitem [{\citenamefont {Ansari}\ \emph {et~al.}(2017)\citenamefont {Ansari},
  \citenamefont {Attarzadeh}, \citenamefont {Nouh},\ and\ \citenamefont
  {Karami}}]{ansari2017application}%
  \BibitemOpen
  \bibfield  {author} {\bibinfo {author} {\bibfnamefont {M.}~\bibnamefont
  {Ansari}}, \bibinfo {author} {\bibfnamefont {M.}~\bibnamefont {Attarzadeh}},
  \bibinfo {author} {\bibfnamefont {M.}~\bibnamefont {Nouh}},\ and\ \bibinfo
  {author} {\bibfnamefont {M.~A.}\ \bibnamefont {Karami}},\ }\bibfield  {title}
  {\bibinfo {title} {Application of magnetoelastic materials in
  spatiotemporally modulated phononic crystals for nonreciprocal wave
  propagation},\ }\href@noop {} {\bibfield  {journal} {\bibinfo  {journal}
  {Smart Mater. Struct.}\ }\textbf {\bibinfo {volume} {27}},\ \bibinfo {pages}
  {015030} (\bibinfo {year} {2017})}\BibitemShut {NoStop}%
\bibitem [{\citenamefont {Huang}\ and\ \citenamefont
  {Zhou}(2019)}]{huang2019mass}%
  \BibitemOpen
  \bibfield  {author} {\bibinfo {author} {\bibfnamefont {J.}~\bibnamefont
  {Huang}}\ and\ \bibinfo {author} {\bibfnamefont {X.}~\bibnamefont {Zhou}},\
  }\bibfield  {title} {\bibinfo {title} {A time-varying mass metamaterial for
  non-reciprocal wave propagation},\ }\href@noop {} {\bibfield  {journal}
  {\bibinfo  {journal} {Int. J. Solids Struct.}\ }\textbf {\bibinfo {volume}
  {164}},\ \bibinfo {pages} {25} (\bibinfo {year} {2019})}\BibitemShut
  {NoStop}%
\bibitem [{\citenamefont {Huang}\ and\ \citenamefont
  {Zhou}(2020)}]{huang2020mass2}%
  \BibitemOpen
  \bibfield  {author} {\bibinfo {author} {\bibfnamefont {J.}~\bibnamefont
  {Huang}}\ and\ \bibinfo {author} {\bibfnamefont {X.}~\bibnamefont {Zhou}},\
  }\bibfield  {title} {\bibinfo {title} {Non-reciprocal metamaterials with
  simultaneously time-varying stiffness and mass},\ }\href@noop {} {\bibfield
  {journal} {\bibinfo  {journal} {J. Appl. Mech.}\ }\textbf {\bibinfo {volume}
  {87}},\ \bibinfo {pages} {071003} (\bibinfo {year} {2020})}\BibitemShut
  {NoStop}%
\bibitem [{\citenamefont {Attarzadeh}\ \emph {et~al.}(2019)\citenamefont
  {Attarzadeh}, \citenamefont {Maleki}, \citenamefont {Crassidis},\ and\
  \citenamefont {Nouh}}]{attarzadeh2019gyro}%
  \BibitemOpen
  \bibfield  {author} {\bibinfo {author} {\bibfnamefont {M.~A.}\ \bibnamefont
  {Attarzadeh}}, \bibinfo {author} {\bibfnamefont {S.}~\bibnamefont {Maleki}},
  \bibinfo {author} {\bibfnamefont {J.}~\bibnamefont {Crassidis}},\ and\
  \bibinfo {author} {\bibfnamefont {M.}~\bibnamefont {Nouh}},\ }\bibfield
  {title} {\bibinfo {title} {Non-reciprocal wave phenomena in energy
  self-reliant gyric metamaterials},\ }\href@noop {} {\bibfield  {journal}
  {\bibinfo  {journal} {J. Acoust. Soc. Am.}\ }\textbf {\bibinfo {volume}
  {146}},\ \bibinfo {pages} {789} (\bibinfo {year} {2019})}\BibitemShut
  {NoStop}%
\bibitem [{\citenamefont {Riva}(2022)}]{Riva_non_Her}%
  \BibitemOpen
  \bibfield  {author} {\bibinfo {author} {\bibfnamefont {E.}~\bibnamefont
  {Riva}},\ }\bibfield  {title} {\bibinfo {title} {Harnessing
  $\mathcal{PT}$-symmetry in non-hermitian stiffness-modulated waveguides},\
  }\href@noop {} {\bibfield  {journal} {\bibinfo  {journal} {Phys. Rev. B}\
  }\textbf {\bibinfo {volume} {105}},\ \bibinfo {pages} {224314} (\bibinfo
  {year} {2022})}\BibitemShut {NoStop}%
\bibitem [{\citenamefont {Bender}\ and\ \citenamefont
  {Boettcher}(1998)}]{bender1998real}%
  \BibitemOpen
  \bibfield  {author} {\bibinfo {author} {\bibfnamefont {C.~M.}\ \bibnamefont
  {Bender}}\ and\ \bibinfo {author} {\bibfnamefont {S.}~\bibnamefont
  {Boettcher}},\ }\bibfield  {title} {\bibinfo {title} {Real spectra in
  non-hermitian hamiltonians having pt symmetry},\ }\href@noop {} {\bibfield
  {journal} {\bibinfo  {journal} {Phys. Rev. Lett.}\ }\textbf {\bibinfo
  {volume} {80}},\ \bibinfo {pages} {5243} (\bibinfo {year}
  {1998})}\BibitemShut {NoStop}%
\bibitem [{\citenamefont {Wang}\ and\ \citenamefont
  {Amirkhizi}(2022)}]{wang2022exceptional}%
  \BibitemOpen
  \bibfield  {author} {\bibinfo {author} {\bibfnamefont {W.}~\bibnamefont
  {Wang}}\ and\ \bibinfo {author} {\bibfnamefont {A.~V.}\ \bibnamefont
  {Amirkhizi}},\ }\bibfield  {title} {\bibinfo {title} {Exceptional points and
  scattering of discrete mechanical metamaterials},\ }\href@noop {} {\bibfield
  {journal} {\bibinfo  {journal} {Eur. Phys. J. Plus}\ }\textbf {\bibinfo
  {volume} {137}},\ \bibinfo {pages} {1} (\bibinfo {year} {2022})}\BibitemShut
  {NoStop}%
\bibitem [{\citenamefont {{\"O}zdemir}\ \emph {et~al.}(2019)\citenamefont
  {{\"O}zdemir}, \citenamefont {Rotter}, \citenamefont {Nori},\ and\
  \citenamefont {Yang}}]{ozdemir2019parity}%
  \BibitemOpen
  \bibfield  {author} {\bibinfo {author} {\bibfnamefont {{\c{S}}.~K.}\
  \bibnamefont {{\"O}zdemir}}, \bibinfo {author} {\bibfnamefont
  {S.}~\bibnamefont {Rotter}}, \bibinfo {author} {\bibfnamefont
  {F.}~\bibnamefont {Nori}},\ and\ \bibinfo {author} {\bibfnamefont
  {L.}~\bibnamefont {Yang}},\ }\bibfield  {title} {\bibinfo {title}
  {Parity--time symmetry and exceptional points in photonics},\ }\href@noop {}
  {\bibfield  {journal} {\bibinfo  {journal} {Nat. Mater.}\ }\textbf {\bibinfo
  {volume} {18}},\ \bibinfo {pages} {783} (\bibinfo {year} {2019})}\BibitemShut
  {NoStop}%
\bibitem [{\citenamefont {Makris}\ \emph {et~al.}(2010)\citenamefont {Makris},
  \citenamefont {El-Ganainy}, \citenamefont {Christodoulides},\ and\
  \citenamefont {Musslimani}}]{makris2010pt}%
  \BibitemOpen
  \bibfield  {author} {\bibinfo {author} {\bibfnamefont {K.~G.}\ \bibnamefont
  {Makris}}, \bibinfo {author} {\bibfnamefont {R.}~\bibnamefont {El-Ganainy}},
  \bibinfo {author} {\bibfnamefont {D.~N.}\ \bibnamefont {Christodoulides}},\
  and\ \bibinfo {author} {\bibfnamefont {Z.~H.}\ \bibnamefont {Musslimani}},\
  }\bibfield  {title} {\bibinfo {title} {Pt-symmetric optical lattices},\
  }\href@noop {} {\bibfield  {journal} {\bibinfo  {journal} {Phys. Rev. A}\
  }\textbf {\bibinfo {volume} {81}},\ \bibinfo {pages} {063807} (\bibinfo
  {year} {2010})}\BibitemShut {NoStop}%
\bibitem [{\citenamefont {Miri}\ \emph {et~al.}(2012)\citenamefont {Miri},
  \citenamefont {Aceves}, \citenamefont {Kottos}, \citenamefont {Kovanis},\
  and\ \citenamefont {Christodoulides}}]{miri2012bragg}%
  \BibitemOpen
  \bibfield  {author} {\bibinfo {author} {\bibfnamefont {M.}~\bibnamefont
  {Miri}}, \bibinfo {author} {\bibfnamefont {A.~B.}\ \bibnamefont {Aceves}},
  \bibinfo {author} {\bibfnamefont {T.}~\bibnamefont {Kottos}}, \bibinfo
  {author} {\bibfnamefont {V.}~\bibnamefont {Kovanis}},\ and\ \bibinfo {author}
  {\bibfnamefont {D.~N.}\ \bibnamefont {Christodoulides}},\ }\bibfield  {title}
  {\bibinfo {title} {Bragg solitons in nonlinear pt-symmetric periodic
  potentials},\ }\href@noop {} {\bibfield  {journal} {\bibinfo  {journal}
  {Phys. Rev. A}\ }\textbf {\bibinfo {volume} {86}},\ \bibinfo {pages} {033801}
  (\bibinfo {year} {2012})}\BibitemShut {NoStop}%
\bibitem [{\citenamefont {Zheng}(2009)}]{zheng2010non}%
  \BibitemOpen
  \bibfield  {author} {\bibinfo {author} {\bibfnamefont {M.~C.}\ \bibnamefont
  {Zheng}},\ }\emph {\bibinfo {title} {Non-Hermitian Dynamics: Example from
  Disordered Microwave Cavities and Classical Optics}},\ \href@noop {}
  {Master's thesis},\ \bibinfo  {school} {Wesleyan University} (\bibinfo {year}
  {2009})\BibitemShut {NoStop}%
\bibitem [{\citenamefont {Yang}\ \emph {et~al.}(2022)\citenamefont {Yang},
  \citenamefont {Yang}, \citenamefont {Guan}, \citenamefont {Zou},\ and\
  \citenamefont {Cheng}}]{yang2022design}%
  \BibitemOpen
  \bibfield  {author} {\bibinfo {author} {\bibfnamefont {W.}~\bibnamefont
  {Yang}}, \bibinfo {author} {\bibfnamefont {Z.}~\bibnamefont {Yang}}, \bibinfo
  {author} {\bibfnamefont {A.}~\bibnamefont {Guan}}, \bibinfo {author}
  {\bibfnamefont {X.}~\bibnamefont {Zou}},\ and\ \bibinfo {author}
  {\bibfnamefont {J.}~\bibnamefont {Cheng}},\ }\bibfield  {title} {\bibinfo
  {title} {Design and experimental demonstration of effective acoustic gain
  medium for pt-symmetric refractive index},\ }\href@noop {} {\bibfield
  {journal} {\bibinfo  {journal} {Appl. Phys. Lett.}\ }\textbf {\bibinfo
  {volume} {120}},\ \bibinfo {pages} {063503} (\bibinfo {year}
  {2022})}\BibitemShut {NoStop}%
\bibitem [{\citenamefont {Gu}\ \emph {et~al.}(2021)\citenamefont {Gu},
  \citenamefont {Gao}, \citenamefont {Cao}, \citenamefont {Liu}, \citenamefont
  {Zhu},\ and\ \citenamefont {Zhu}}]{gu2021controlling}%
  \BibitemOpen
  \bibfield  {author} {\bibinfo {author} {\bibfnamefont {Z.}~\bibnamefont
  {Gu}}, \bibinfo {author} {\bibfnamefont {H.}~\bibnamefont {Gao}}, \bibinfo
  {author} {\bibfnamefont {P.}~\bibnamefont {Cao}}, \bibinfo {author}
  {\bibfnamefont {T.}~\bibnamefont {Liu}}, \bibinfo {author} {\bibfnamefont
  {X.}~\bibnamefont {Zhu}},\ and\ \bibinfo {author} {\bibfnamefont
  {J.}~\bibnamefont {Zhu}},\ }\bibfield  {title} {\bibinfo {title} {Controlling
  sound in non-hermitian acoustic systems},\ }\href@noop {} {\bibfield
  {journal} {\bibinfo  {journal} {Phys. Rev. Appl.}\ }\textbf {\bibinfo
  {volume} {16}},\ \bibinfo {pages} {057001} (\bibinfo {year}
  {2021})}\BibitemShut {NoStop}%
\bibitem [{\citenamefont {Zhu}\ \emph {et~al.}(2014)\citenamefont {Zhu},
  \citenamefont {Ramezani}, \citenamefont {Shi}, \citenamefont {Zhu},\ and\
  \citenamefont {Zhang}}]{zhu2014p}%
  \BibitemOpen
  \bibfield  {author} {\bibinfo {author} {\bibfnamefont {X.}~\bibnamefont
  {Zhu}}, \bibinfo {author} {\bibfnamefont {H.}~\bibnamefont {Ramezani}},
  \bibinfo {author} {\bibfnamefont {C.}~\bibnamefont {Shi}}, \bibinfo {author}
  {\bibfnamefont {J.}~\bibnamefont {Zhu}},\ and\ \bibinfo {author}
  {\bibfnamefont {X.}~\bibnamefont {Zhang}},\ }\bibfield  {title} {\bibinfo
  {title} {P t-symmetric acoustics},\ }\href@noop {} {\bibfield  {journal}
  {\bibinfo  {journal} {Phys. Rev. X}\ }\textbf {\bibinfo {volume} {4}},\
  \bibinfo {pages} {031042} (\bibinfo {year} {2014})}\BibitemShut {NoStop}%
\bibitem [{\citenamefont {Geng}\ \emph {et~al.}(2021)\citenamefont {Geng},
  \citenamefont {Zhang}, \citenamefont {Zhang},\ and\ \citenamefont
  {Zhou}}]{geng2021chiral}%
  \BibitemOpen
  \bibfield  {author} {\bibinfo {author} {\bibfnamefont {L.}~\bibnamefont
  {Geng}}, \bibinfo {author} {\bibfnamefont {W.}~\bibnamefont {Zhang}},
  \bibinfo {author} {\bibfnamefont {X.}~\bibnamefont {Zhang}},\ and\ \bibinfo
  {author} {\bibfnamefont {X.}~\bibnamefont {Zhou}},\ }\bibfield  {title}
  {\bibinfo {title} {Chiral mode transfer of symmetry-broken states in
  anti-parity-time-symmetric mechanical system},\ }\href@noop {} {\bibfield
  {journal} {\bibinfo  {journal} {Proc. Royal Soc. A}\ }\textbf {\bibinfo
  {volume} {477}},\ \bibinfo {pages} {20210641} (\bibinfo {year}
  {2021})}\BibitemShut {NoStop}%
\bibitem [{\citenamefont {Christensen}\ \emph {et~al.}(2016)\citenamefont
  {Christensen}, \citenamefont {Willatzen}, \citenamefont {Velasco},\ and\
  \citenamefont {Lu}}]{christensen2016parity}%
  \BibitemOpen
  \bibfield  {author} {\bibinfo {author} {\bibfnamefont {J.}~\bibnamefont
  {Christensen}}, \bibinfo {author} {\bibfnamefont {M.}~\bibnamefont
  {Willatzen}}, \bibinfo {author} {\bibfnamefont {V.}~\bibnamefont {Velasco}},\
  and\ \bibinfo {author} {\bibfnamefont {M.-H.}\ \bibnamefont {Lu}},\
  }\bibfield  {title} {\bibinfo {title} {Parity-time synthetic phononic
  media},\ }\href@noop {} {\bibfield  {journal} {\bibinfo  {journal} {Phys.
  Rev. Lett.}\ }\textbf {\bibinfo {volume} {116}},\ \bibinfo {pages} {207601}
  (\bibinfo {year} {2016})}\BibitemShut {NoStop}%
\bibitem [{\citenamefont {Braghini}\ \emph {et~al.}(2021)\citenamefont
  {Braghini}, \citenamefont {Villani}, \citenamefont {Rosa},\ and\
  \citenamefont {de~F~Arruda}}]{braghini2021non}%
  \BibitemOpen
  \bibfield  {author} {\bibinfo {author} {\bibfnamefont {D.}~\bibnamefont
  {Braghini}}, \bibinfo {author} {\bibfnamefont {L.~G.}\ \bibnamefont
  {Villani}}, \bibinfo {author} {\bibfnamefont {M.~I.}\ \bibnamefont {Rosa}},\
  and\ \bibinfo {author} {\bibfnamefont {J.~R.}\ \bibnamefont {de~F~Arruda}},\
  }\bibfield  {title} {\bibinfo {title} {Non-hermitian elastic waveguides with
  piezoelectric feedback actuation: non-reciprocal bands and skin modes},\
  }\href@noop {} {\bibfield  {journal} {\bibinfo  {journal} {J. Phys. D: Appl.
  Phys.}\ }\textbf {\bibinfo {volume} {54}},\ \bibinfo {pages} {285302}
  (\bibinfo {year} {2021})}\BibitemShut {NoStop}%
\bibitem [{\citenamefont {Feng}\ \emph {et~al.}(2013)\citenamefont {Feng},
  \citenamefont {Xu}, \citenamefont {Fegadolli}, \citenamefont {Lu},
  \citenamefont {Oliveira}, \citenamefont {Almeida}, \citenamefont {Chen},\
  and\ \citenamefont {Scherer}}]{feng2013experimental}%
  \BibitemOpen
  \bibfield  {author} {\bibinfo {author} {\bibfnamefont {L.}~\bibnamefont
  {Feng}}, \bibinfo {author} {\bibfnamefont {Y.}~\bibnamefont {Xu}}, \bibinfo
  {author} {\bibfnamefont {W.~S.}\ \bibnamefont {Fegadolli}}, \bibinfo {author}
  {\bibfnamefont {M.}~\bibnamefont {Lu}}, \bibinfo {author} {\bibfnamefont
  {J.~E.}\ \bibnamefont {Oliveira}}, \bibinfo {author} {\bibfnamefont {V.~R.}\
  \bibnamefont {Almeida}}, \bibinfo {author} {\bibfnamefont {Y.}~\bibnamefont
  {Chen}},\ and\ \bibinfo {author} {\bibfnamefont {A.}~\bibnamefont
  {Scherer}},\ }\bibfield  {title} {\bibinfo {title} {Experimental
  demonstration of a unidirectional reflectionless parity-time metamaterial at
  optical frequencies},\ }\href@noop {} {\bibfield  {journal} {\bibinfo
  {journal} {Nat. Mater.}\ }\textbf {\bibinfo {volume} {12}},\ \bibinfo {pages}
  {108} (\bibinfo {year} {2013})}\BibitemShut {NoStop}%
\bibitem [{\citenamefont {Ge}\ \emph {et~al.}(2011)\citenamefont {Ge},
  \citenamefont {Chong}, \citenamefont {Rotter}, \citenamefont {T{\"u}reci},\
  and\ \citenamefont {Stone}}]{ge2011unconventional}%
  \BibitemOpen
  \bibfield  {author} {\bibinfo {author} {\bibfnamefont {L.}~\bibnamefont
  {Ge}}, \bibinfo {author} {\bibfnamefont {Y.}~\bibnamefont {Chong}}, \bibinfo
  {author} {\bibfnamefont {S.}~\bibnamefont {Rotter}}, \bibinfo {author}
  {\bibfnamefont {H.~E.}\ \bibnamefont {T{\"u}reci}},\ and\ \bibinfo {author}
  {\bibfnamefont {A.~D.}\ \bibnamefont {Stone}},\ }\bibfield  {title} {\bibinfo
  {title} {Unconventional modes in lasers with spatially varying gain and
  loss},\ }\href@noop {} {\bibfield  {journal} {\bibinfo  {journal} {Phys. Rev.
  A}\ }\textbf {\bibinfo {volume} {84}},\ \bibinfo {pages} {023820} (\bibinfo
  {year} {2011})}\BibitemShut {NoStop}%
\bibitem [{\citenamefont {Hodaei}\ \emph {et~al.}(2014)\citenamefont {Hodaei},
  \citenamefont {Miri}, \citenamefont {Heinrich}, \citenamefont
  {Christodoulides},\ and\ \citenamefont {Khajavikhan}}]{hodaei2014parity}%
  \BibitemOpen
  \bibfield  {author} {\bibinfo {author} {\bibfnamefont {H.}~\bibnamefont
  {Hodaei}}, \bibinfo {author} {\bibfnamefont {M.}~\bibnamefont {Miri}},
  \bibinfo {author} {\bibfnamefont {M.}~\bibnamefont {Heinrich}}, \bibinfo
  {author} {\bibfnamefont {D.~N.}\ \bibnamefont {Christodoulides}},\ and\
  \bibinfo {author} {\bibfnamefont {M.}~\bibnamefont {Khajavikhan}},\
  }\bibfield  {title} {\bibinfo {title} {Parity-time--symmetric microring
  lasers},\ }\href@noop {} {\bibfield  {journal} {\bibinfo  {journal}
  {Science}\ }\textbf {\bibinfo {volume} {346}},\ \bibinfo {pages} {975}
  (\bibinfo {year} {2014})}\BibitemShut {NoStop}%
\bibitem [{\citenamefont {Miroshnichenko}\ \emph {et~al.}(2011)\citenamefont
  {Miroshnichenko}, \citenamefont {Malomed},\ and\ \citenamefont
  {Kivshar}}]{miroshnichenko2011nonlinearly}%
  \BibitemOpen
  \bibfield  {author} {\bibinfo {author} {\bibfnamefont {A.~E.}\ \bibnamefont
  {Miroshnichenko}}, \bibinfo {author} {\bibfnamefont {B.~A.}\ \bibnamefont
  {Malomed}},\ and\ \bibinfo {author} {\bibfnamefont {Y.~S.}\ \bibnamefont
  {Kivshar}},\ }\bibfield  {title} {\bibinfo {title} {Nonlinearly pt-symmetric
  systems: Spontaneous symmetry breaking and transmission resonances},\
  }\href@noop {} {\bibfield  {journal} {\bibinfo  {journal} {Phys. Rev. A}\
  }\textbf {\bibinfo {volume} {84}},\ \bibinfo {pages} {012123} (\bibinfo
  {year} {2011})}\BibitemShut {NoStop}%
\bibitem [{\citenamefont {R{\"u}ter}\ \emph {et~al.}(2010)\citenamefont
  {R{\"u}ter}, \citenamefont {Makris}, \citenamefont {El-Ganainy},
  \citenamefont {Christodoulides}, \citenamefont {Segev},\ and\ \citenamefont
  {Kip}}]{ruter2010observation}%
  \BibitemOpen
  \bibfield  {author} {\bibinfo {author} {\bibfnamefont {C.~E.}\ \bibnamefont
  {R{\"u}ter}}, \bibinfo {author} {\bibfnamefont {K.~G.}\ \bibnamefont
  {Makris}}, \bibinfo {author} {\bibfnamefont {R.}~\bibnamefont {El-Ganainy}},
  \bibinfo {author} {\bibfnamefont {D.~N.}\ \bibnamefont {Christodoulides}},
  \bibinfo {author} {\bibfnamefont {M.}~\bibnamefont {Segev}},\ and\ \bibinfo
  {author} {\bibfnamefont {D.}~\bibnamefont {Kip}},\ }\bibfield  {title}
  {\bibinfo {title} {Observation of parity--time symmetry in optics},\
  }\href@noop {} {\bibfield  {journal} {\bibinfo  {journal} {Nat. Phys.}\
  }\textbf {\bibinfo {volume} {6}},\ \bibinfo {pages} {192} (\bibinfo {year}
  {2010})}\BibitemShut {NoStop}%
\bibitem [{\citenamefont {Chang}\ \emph {et~al.}(2014)\citenamefont {Chang},
  \citenamefont {Jiang}, \citenamefont {Hua}, \citenamefont {Yang},
  \citenamefont {Wen}, \citenamefont {Jiang}, \citenamefont {Li}, \citenamefont
  {Wang},\ and\ \citenamefont {Xiao}}]{chang2014parity}%
  \BibitemOpen
  \bibfield  {author} {\bibinfo {author} {\bibfnamefont {L.}~\bibnamefont
  {Chang}}, \bibinfo {author} {\bibfnamefont {X.}~\bibnamefont {Jiang}},
  \bibinfo {author} {\bibfnamefont {S.}~\bibnamefont {Hua}}, \bibinfo {author}
  {\bibfnamefont {C.}~\bibnamefont {Yang}}, \bibinfo {author} {\bibfnamefont
  {J.}~\bibnamefont {Wen}}, \bibinfo {author} {\bibfnamefont {L.}~\bibnamefont
  {Jiang}}, \bibinfo {author} {\bibfnamefont {G.}~\bibnamefont {Li}}, \bibinfo
  {author} {\bibfnamefont {G.}~\bibnamefont {Wang}},\ and\ \bibinfo {author}
  {\bibfnamefont {M.}~\bibnamefont {Xiao}},\ }\bibfield  {title} {\bibinfo
  {title} {Parity--time symmetry and variable optical isolation in
  active--passive-coupled microresonators},\ }\href@noop {} {\bibfield
  {journal} {\bibinfo  {journal} {Nat. Photon.}\ }\textbf {\bibinfo {volume}
  {8}},\ \bibinfo {pages} {524} (\bibinfo {year} {2014})}\BibitemShut {NoStop}%
\bibitem [{\citenamefont {Fleury}\ \emph {et~al.}(2015)\citenamefont {Fleury},
  \citenamefont {Sounas},\ and\ \citenamefont {Alu}}]{fleury2015invisible}%
  \BibitemOpen
  \bibfield  {author} {\bibinfo {author} {\bibfnamefont {R.}~\bibnamefont
  {Fleury}}, \bibinfo {author} {\bibfnamefont {D.}~\bibnamefont {Sounas}},\
  and\ \bibinfo {author} {\bibfnamefont {A.}~\bibnamefont {Alu}},\ }\bibfield
  {title} {\bibinfo {title} {An invisible acoustic sensor based on parity-time
  symmetry},\ }\href@noop {} {\bibfield  {journal} {\bibinfo  {journal} {Nat.
  Commun.}\ }\textbf {\bibinfo {volume} {6}},\ \bibinfo {pages} {1} (\bibinfo
  {year} {2015})}\BibitemShut {NoStop}%
\bibitem [{\citenamefont {Li}\ \emph {et~al.}(2020)\citenamefont {Li},
  \citenamefont {Moussa}, \citenamefont {Sounas},\ and\ \citenamefont
  {Al{\`u}}}]{li2020parity}%
  \BibitemOpen
  \bibfield  {author} {\bibinfo {author} {\bibfnamefont {H.}~\bibnamefont
  {Li}}, \bibinfo {author} {\bibfnamefont {H.}~\bibnamefont {Moussa}}, \bibinfo
  {author} {\bibfnamefont {D.}~\bibnamefont {Sounas}},\ and\ \bibinfo {author}
  {\bibfnamefont {A.}~\bibnamefont {Al{\`u}}},\ }\bibfield  {title} {\bibinfo
  {title} {Parity-time symmetry based on time modulation},\ }\href@noop {}
  {\bibfield  {journal} {\bibinfo  {journal} {Phys. Rev. Appl.}\ }\textbf
  {\bibinfo {volume} {14}},\ \bibinfo {pages} {031002} (\bibinfo {year}
  {2020})}\BibitemShut {NoStop}%
\bibitem [{\citenamefont {Song}\ \emph {et~al.}(2019)\citenamefont {Song},
  \citenamefont {Shi}, \citenamefont {Lin},\ and\ \citenamefont
  {Fan}}]{song2019direction}%
  \BibitemOpen
  \bibfield  {author} {\bibinfo {author} {\bibfnamefont {A.~Y.}\ \bibnamefont
  {Song}}, \bibinfo {author} {\bibfnamefont {Y.}~\bibnamefont {Shi}}, \bibinfo
  {author} {\bibfnamefont {Q.}~\bibnamefont {Lin}},\ and\ \bibinfo {author}
  {\bibfnamefont {S.}~\bibnamefont {Fan}},\ }\bibfield  {title} {\bibinfo
  {title} {Direction-dependent parity-time phase transition and nonreciprocal
  amplification with dynamic gain-loss modulation},\ }\href@noop {} {\bibfield
  {journal} {\bibinfo  {journal} {Phys. Rev. A}\ }\textbf {\bibinfo {volume}
  {99}},\ \bibinfo {pages} {013824} (\bibinfo {year} {2019})}\BibitemShut
  {NoStop}%
\bibitem [{\citenamefont {Liu}\ \emph {et~al.}(2020)\citenamefont {Liu},
  \citenamefont {Qin}, \citenamefont {Wang},\ and\ \citenamefont
  {Lu}}]{liu2020scattering}%
  \BibitemOpen
  \bibfield  {author} {\bibinfo {author} {\bibfnamefont {Q.}~\bibnamefont
  {Liu}}, \bibinfo {author} {\bibfnamefont {C.}~\bibnamefont {Qin}}, \bibinfo
  {author} {\bibfnamefont {B.}~\bibnamefont {Wang}},\ and\ \bibinfo {author}
  {\bibfnamefont {P.}~\bibnamefont {Lu}},\ }\bibfield  {title} {\bibinfo
  {title} {Scattering singularities of optical waveguides under complex
  modulation},\ }\href@noop {} {\bibfield  {journal} {\bibinfo  {journal}
  {Phys. Rev. A}\ }\textbf {\bibinfo {volume} {101}},\ \bibinfo {pages}
  {033818} (\bibinfo {year} {2020})}\BibitemShut {NoStop}%
\bibitem [{\citenamefont {Wang}\ \emph {et~al.}(2018)\citenamefont {Wang},
  \citenamefont {Zhang},\ and\ \citenamefont {Chan}}]{wang2018photonic}%
  \BibitemOpen
  \bibfield  {author} {\bibinfo {author} {\bibfnamefont {N.}~\bibnamefont
  {Wang}}, \bibinfo {author} {\bibfnamefont {Z.}~\bibnamefont {Zhang}},\ and\
  \bibinfo {author} {\bibfnamefont {C.~T.}\ \bibnamefont {Chan}},\ }\bibfield
  {title} {\bibinfo {title} {Photonic floquet media with a complex
  time-periodic permittivity},\ }\href@noop {} {\bibfield  {journal} {\bibinfo
  {journal} {Phys. Rev. B}\ }\textbf {\bibinfo {volume} {98}},\ \bibinfo
  {pages} {085142} (\bibinfo {year} {2018})}\BibitemShut {NoStop}%
\bibitem [{\citenamefont {Koutserimpas}\ \emph {et~al.}(2018)\citenamefont
  {Koutserimpas}, \citenamefont {Al{\`u}},\ and\ \citenamefont
  {Fleury}}]{koutserimpas2018parametric}%
  \BibitemOpen
  \bibfield  {author} {\bibinfo {author} {\bibfnamefont {T.~T.}\ \bibnamefont
  {Koutserimpas}}, \bibinfo {author} {\bibfnamefont {A.}~\bibnamefont
  {Al{\`u}}},\ and\ \bibinfo {author} {\bibfnamefont {R.}~\bibnamefont
  {Fleury}},\ }\bibfield  {title} {\bibinfo {title} {Parametric amplification
  and bidirectional invisibility in pt-symmetric time-floquet systems},\
  }\href@noop {} {\bibfield  {journal} {\bibinfo  {journal} {Phys. Rev. A}\
  }\textbf {\bibinfo {volume} {97}},\ \bibinfo {pages} {013839} (\bibinfo
  {year} {2018})}\BibitemShut {NoStop}%
\bibitem [{\citenamefont {Wu}\ \emph {et~al.}(2019)\citenamefont {Wu},
  \citenamefont {Chen},\ and\ \citenamefont {Huang}}]{wu2019asymmetric}%
  \BibitemOpen
  \bibfield  {author} {\bibinfo {author} {\bibfnamefont {Q.}~\bibnamefont
  {Wu}}, \bibinfo {author} {\bibfnamefont {Y.}~\bibnamefont {Chen}},\ and\
  \bibinfo {author} {\bibfnamefont {G.}~\bibnamefont {Huang}},\ }\bibfield
  {title} {\bibinfo {title} {Asymmetric scattering of flexural waves in a
  parity-time symmetric metamaterial beam},\ }\href@noop {} {\bibfield
  {journal} {\bibinfo  {journal} {J. Acoust. Soc. Am.}\ }\textbf {\bibinfo
  {volume} {146}},\ \bibinfo {pages} {850} (\bibinfo {year}
  {2019})}\BibitemShut {NoStop}%
\bibitem [{\citenamefont {Park}\ and\ \citenamefont
  {Min}(2021)}]{park2021spatiotemporal}%
  \BibitemOpen
  \bibfield  {author} {\bibinfo {author} {\bibfnamefont {J.}~\bibnamefont
  {Park}}\ and\ \bibinfo {author} {\bibfnamefont {B.}~\bibnamefont {Min}},\
  }\bibfield  {title} {\bibinfo {title} {Spatiotemporal plane wave expansion
  method for arbitrary space--time periodic photonic media},\ }\href@noop {}
  {\bibfield  {journal} {\bibinfo  {journal} {Opt. Lett.}\ }\textbf {\bibinfo
  {volume} {46}},\ \bibinfo {pages} {484} (\bibinfo {year} {2021})}\BibitemShut
  {NoStop}%
\bibitem [{\citenamefont {Khelif}\ \emph {et~al.}(2006)\citenamefont {Khelif},
  \citenamefont {Aoubiza}, \citenamefont {Mohammadi}, \citenamefont {Adibi},\
  and\ \citenamefont {Laude}}]{khelif2006complete}%
  \BibitemOpen
  \bibfield  {author} {\bibinfo {author} {\bibfnamefont {A.}~\bibnamefont
  {Khelif}}, \bibinfo {author} {\bibfnamefont {B.}~\bibnamefont {Aoubiza}},
  \bibinfo {author} {\bibfnamefont {S.}~\bibnamefont {Mohammadi}}, \bibinfo
  {author} {\bibfnamefont {A.}~\bibnamefont {Adibi}},\ and\ \bibinfo {author}
  {\bibfnamefont {V.}~\bibnamefont {Laude}},\ }\bibfield  {title} {\bibinfo
  {title} {Complete band gaps in two-dimensional phononic crystal slabs},\
  }\href@noop {} {\bibfield  {journal} {\bibinfo  {journal} {Phys. Rev. E}\
  }\textbf {\bibinfo {volume} {74}},\ \bibinfo {pages} {046610} (\bibinfo
  {year} {2006})}\BibitemShut {NoStop}%
\bibitem [{\citenamefont {Hsu}\ and\ \citenamefont
  {Wu}(2006)}]{hsu2006efficient}%
  \BibitemOpen
  \bibfield  {author} {\bibinfo {author} {\bibfnamefont {J.}~\bibnamefont
  {Hsu}}\ and\ \bibinfo {author} {\bibfnamefont {T.}~\bibnamefont {Wu}},\
  }\bibfield  {title} {\bibinfo {title} {Efficient formulation for
  band-structure calculations of two-dimensional phononic-crystal plates},\
  }\href@noop {} {\bibfield  {journal} {\bibinfo  {journal} {Phys. Rev. B}\
  }\textbf {\bibinfo {volume} {74}},\ \bibinfo {pages} {144303} (\bibinfo
  {year} {2006})}\BibitemShut {NoStop}%
\bibitem [{\citenamefont {Laude}\ \emph {et~al.}(2009)\citenamefont {Laude},
  \citenamefont {Achaoui}, \citenamefont {Benchabane},\ and\ \citenamefont
  {Khelif}}]{laude2009evanescent}%
  \BibitemOpen
  \bibfield  {author} {\bibinfo {author} {\bibfnamefont {V.}~\bibnamefont
  {Laude}}, \bibinfo {author} {\bibfnamefont {Y.}~\bibnamefont {Achaoui}},
  \bibinfo {author} {\bibfnamefont {S.}~\bibnamefont {Benchabane}},\ and\
  \bibinfo {author} {\bibfnamefont {A.}~\bibnamefont {Khelif}},\ }\bibfield
  {title} {\bibinfo {title} {Evanescent bloch waves and the complex band
  structure of phononic crystals},\ }\href@noop {} {\bibfield  {journal}
  {\bibinfo  {journal} {Phys. Rev. B}\ }\textbf {\bibinfo {volume} {80}},\
  \bibinfo {pages} {092301} (\bibinfo {year} {2009})}\BibitemShut {NoStop}%
\bibitem [{\citenamefont {Al~Ba'ba'a}\ \emph {et~al.}(2017)\citenamefont
  {Al~Ba'ba'a}, \citenamefont {Nouh},\ and\ \citenamefont
  {Singh}}]{al2017pole}%
  \BibitemOpen
  \bibfield  {author} {\bibinfo {author} {\bibfnamefont {H.}~\bibnamefont
  {Al~Ba'ba'a}}, \bibinfo {author} {\bibfnamefont {M.}~\bibnamefont {Nouh}},\
  and\ \bibinfo {author} {\bibfnamefont {T.}~\bibnamefont {Singh}},\ }\bibfield
   {title} {\bibinfo {title} {Pole distribution in finite phononic crystals:
  Understanding bragg-effects through closed-form system dynamics},\
  }\href@noop {} {\bibfield  {journal} {\bibinfo  {journal} {J. Acoust. Soc.
  Am.}\ }\textbf {\bibinfo {volume} {142}},\ \bibinfo {pages} {1399} (\bibinfo
  {year} {2017})}\BibitemShut {NoStop}%
\bibitem [{\citenamefont {Al~Ba'ba'a}\ and\ \citenamefont
  {Nouh}(2017)}]{Nouh2017}%
  \BibitemOpen
  \bibfield  {author} {\bibinfo {author} {\bibfnamefont {H.}~\bibnamefont
  {Al~Ba'ba'a}}\ and\ \bibinfo {author} {\bibfnamefont {M.}~\bibnamefont
  {Nouh}},\ }\bibfield  {title} {\bibinfo {title} {An investigation of
  vibrational power flow in one-dimensional dissipative phononic structures},\
  }\href@noop {} {\bibfield  {journal} {\bibinfo  {journal} {J. Vib. Acoust.}\
  }\textbf {\bibinfo {volume} {139}},\ \bibinfo {pages} {021003} (\bibinfo
  {year} {2017})}\BibitemShut {NoStop}%
\bibitem [{\citenamefont {Bastawrous}\ and\ \citenamefont
  {Hussein}(2022)}]{bastawrous2022closed}%
  \BibitemOpen
  \bibfield  {author} {\bibinfo {author} {\bibfnamefont {M.~V.}\ \bibnamefont
  {Bastawrous}}\ and\ \bibinfo {author} {\bibfnamefont {M.~I.}\ \bibnamefont
  {Hussein}},\ }\bibfield  {title} {\bibinfo {title} {Closed-form existence
  conditions for bandgap resonances in a finite periodic chain under general
  boundary conditions},\ }\href@noop {} {\bibfield  {journal} {\bibinfo
  {journal} {J. Acoust. Soc. Am.}\ }\textbf {\bibinfo {volume} {151}},\
  \bibinfo {pages} {286} (\bibinfo {year} {2022})}\BibitemShut {NoStop}%
\bibitem [{\citenamefont {Tisseur}\ and\ \citenamefont
  {Meerbergen}(2001)}]{tisseur2001quadratic}%
  \BibitemOpen
  \bibfield  {author} {\bibinfo {author} {\bibfnamefont {F.}~\bibnamefont
  {Tisseur}}\ and\ \bibinfo {author} {\bibfnamefont {K.}~\bibnamefont
  {Meerbergen}},\ }\bibfield  {title} {\bibinfo {title} {The quadratic
  eigenvalue problem},\ }\href@noop {} {\bibfield  {journal} {\bibinfo
  {journal} {SIAM review}\ }\textbf {\bibinfo {volume} {43}},\ \bibinfo {pages}
  {235} (\bibinfo {year} {2001})}\BibitemShut {NoStop}%
\bibitem [{\citenamefont {Galiffi}\ \emph {et~al.}(2019)\citenamefont
  {Galiffi}, \citenamefont {Huidobro},\ and\ \citenamefont
  {Pendry}}]{galiffi2019broadband}%
  \BibitemOpen
  \bibfield  {author} {\bibinfo {author} {\bibfnamefont {E.}~\bibnamefont
  {Galiffi}}, \bibinfo {author} {\bibfnamefont {P.}~\bibnamefont {Huidobro}},\
  and\ \bibinfo {author} {\bibfnamefont {J.~B.}\ \bibnamefont {Pendry}},\
  }\bibfield  {title} {\bibinfo {title} {Broadband nonreciprocal amplification
  in luminal metamaterials},\ }\href@noop {} {\bibfield  {journal} {\bibinfo
  {journal} {Phys. Rev. Lett.}\ }\textbf {\bibinfo {volume} {123}},\ \bibinfo
  {pages} {206101} (\bibinfo {year} {2019})}\BibitemShut {NoStop}%
\bibitem [{\citenamefont {Cassedy}(1962)}]{cassedy1962temporal}%
  \BibitemOpen
  \bibfield  {author} {\bibinfo {author} {\bibfnamefont {E.}~\bibnamefont
  {Cassedy}},\ }\bibfield  {title} {\bibinfo {title} {Temporal instabilities in
  traveling-wave parametric amplifiers (correspondence)},\ }\href@noop {}
  {\bibfield  {journal} {\bibinfo  {journal} {IEEE Trans. Microw. Theory
  Tech.}\ }\textbf {\bibinfo {volume} {10}},\ \bibinfo {pages} {86} (\bibinfo
  {year} {1962})}\BibitemShut {NoStop}%
\bibitem [{\citenamefont {Cassedy}(1967)}]{cassedy1967dispersion}%
  \BibitemOpen
  \bibfield  {author} {\bibinfo {author} {\bibfnamefont {E.}~\bibnamefont
  {Cassedy}},\ }\bibfield  {title} {\bibinfo {title} {Dispersion relations in
  time-space periodic media: Part ii, unstable interactions},\ }\href@noop {}
  {\bibfield  {journal} {\bibinfo  {journal} {Proc. IEEE}\ }\textbf {\bibinfo
  {volume} {55}},\ \bibinfo {pages} {1154} (\bibinfo {year}
  {1967})}\BibitemShut {NoStop}%
\bibitem [{\citenamefont {Moghaddaszadeh}\ \emph {et~al.}(2021)\citenamefont
  {Moghaddaszadeh}, \citenamefont {Adlakha}, \citenamefont {Attarzadeh},
  \citenamefont {Aref},\ and\ \citenamefont
  {Nouh}}]{moghaddaszadeh2021nonreciprocal}%
  \BibitemOpen
  \bibfield  {author} {\bibinfo {author} {\bibfnamefont {M.}~\bibnamefont
  {Moghaddaszadeh}}, \bibinfo {author} {\bibfnamefont {R.}~\bibnamefont
  {Adlakha}}, \bibinfo {author} {\bibfnamefont {M.~A.}\ \bibnamefont
  {Attarzadeh}}, \bibinfo {author} {\bibfnamefont {A.}~\bibnamefont {Aref}},\
  and\ \bibinfo {author} {\bibfnamefont {M.}~\bibnamefont {Nouh}},\ }\bibfield
  {title} {\bibinfo {title} {Nonreciprocal elastic wave beaming in dynamic
  phased arrays},\ }\href@noop {} {\bibfield  {journal} {\bibinfo  {journal}
  {Phys. Rev. Appl.}\ }\textbf {\bibinfo {volume} {16}},\ \bibinfo {pages}
  {034033} (\bibinfo {year} {2021})}\BibitemShut {NoStop}%
\bibitem [{\citenamefont {Wiersig}(2020)}]{wiersig2020review}%
  \BibitemOpen
  \bibfield  {author} {\bibinfo {author} {\bibfnamefont {J.}~\bibnamefont
  {Wiersig}},\ }\bibfield  {title} {\bibinfo {title} {Review of exceptional
  point-based sensors},\ }\href@noop {} {\bibfield  {journal} {\bibinfo
  {journal} {Photonics Res.}\ }\textbf {\bibinfo {volume} {8}},\ \bibinfo
  {pages} {1457} (\bibinfo {year} {2020})}\BibitemShut {NoStop}%
\bibitem [{\citenamefont {Rosa}\ \emph {et~al.}(2021)\citenamefont {Rosa},
  \citenamefont {Mazzotti},\ and\ \citenamefont
  {Ruzzene}}]{rosa2021exceptional}%
  \BibitemOpen
  \bibfield  {author} {\bibinfo {author} {\bibfnamefont {M.~I.}\ \bibnamefont
  {Rosa}}, \bibinfo {author} {\bibfnamefont {M.}~\bibnamefont {Mazzotti}},\
  and\ \bibinfo {author} {\bibfnamefont {M.}~\bibnamefont {Ruzzene}},\
  }\bibfield  {title} {\bibinfo {title} {Exceptional points and enhanced
  sensitivity in pt-symmetric continuous elastic media},\ }\href@noop {}
  {\bibfield  {journal} {\bibinfo  {journal} {J. Mech. Phys. Solids}\ }\textbf
  {\bibinfo {volume} {149}},\ \bibinfo {pages} {104325} (\bibinfo {year}
  {2021})}\BibitemShut {NoStop}%
\end{thebibliography}%
\end{document}